\documentclass{pasj00}
\usepackage{xspace}

\begin{document}
%\SetRunningHead{Author(s) in page-head}{Running Head}
\SetRunningHead{Itoh et al.}
{Suzaku wide-band X-ray Spectroscopy of the Seyfert 2 AGN in NGC~4945}
\Received{2007/6/18}%{yyyy/mm/dd}
\Accepted{2007/8/3}%{yyyy/mm/dd}

\title{
  \textrm{Suzaku} wide-band X-ray Spectroscopy of the Seyfert 2 AGN \\
  in NGC~4945
  }

\author{%
  Takeshi \textsc{Itoh}\altaffilmark{1},
  Chris \textsc{Done}\altaffilmark{2},
  Kazuo \textsc{Makishima}\altaffilmark{1,3},
  Grzegorz \textsc{Madejski}\altaffilmark{4},
  Hisamitsu \textsc{Awaki}\altaffilmark{5},\\
  Poshak \textsc{Gandhi}\altaffilmark{3},
  Naoki \textsc{Isobe}\altaffilmark{3},
  Gulab C.  \textsc{Dewangan}\altaffilmark{6},
  Richard E. \textsc{Griffiths}\altaffilmark{6},\\
  Naohisa \textsc{Anabuki}\altaffilmark{7},
  Takashi \textsc{Okajima}\altaffilmark{8,9},
  James \textsc{Reeves}\altaffilmark{8,9}, \\
  Tadayuki \textsc{Takahashi}\altaffilmark{10},
  Yoshihiro \textsc{Ueda}\altaffilmark{11}, 
  Satohi \textsc{Eguchi}\altaffilmark{11}, 
  and
  Tahir \textsc{Yaqoob}\altaffilmark{9,12}}
\altaffiltext{1}{
  Department of Physics, Faculty of Science, University of Tokyo,
  7-3-1 Hongo, Bunkyo-ku, Tokyo, \\113-0033, Japan}
\email{titoh@amalthea.phys.s.u-tokyo.ac.jp}
\altaffiltext{2}{
  Department of Physics, University of Durham, South Rd, DH1 3LE Durham, UK}
\altaffiltext{3}{
  Cosmic Radiation Laboratory, 
  Institute of Physical and Chemical Research (RIKEN)\\
  2-1 Hirosawa, Wako, Saitama, 351-0198, Japan}
\altaffiltext{4}{
  Stanford Linear Accelerator Center, 2575 Sand Hill Road, Menlo Park, 
  CA 94025, USA
  }
\altaffiltext{5}{
  Department of Physics and Astronomy, 
  Ehime University, Matsuyama 790-8577, Japan
  }
\altaffiltext{6}{
  Department of Physics, Carnegie Mellon University, 5000 Forbes Avenue, \\
  Pittsburgh, PA 15213
  }
\altaffiltext{7}{
  Department of Earth and Space Science, 
  Osaka University, 1-1 Machikaneyama, \\
  Toyonaka, Osaka 560-0043
  }
\altaffiltext{8}{
  Exploration of the Universe Division, Code 662, 
  NASA Goddard Space Flight Center,\\
  Greenbelt Road, Greenbelt, MD 20771, USA
}
\altaffiltext{9}{
  Department of Physics and Astronomy, 
  Johns Hopkins University, 3400 N Charles Street, \\
  Baltimore, MD 21218, USA
  }
\altaffiltext{10}{
  Institute of Space and Astronautical Science, JAXA,
  3-1-1, Yoshinodai, Sagamihara, \\Kanagawa, 229-8510, Japan
}
\altaffiltext{11}{
Department of Astronomy, Kyoto University, 
Sakyo-ku, Kyoto 606-8502, Japan
}
\altaffiltext{12}{
Astrophysics Science Division, Code 662, 
NASA Goddard Space Flight Center, \\
Greenbelt, MD 20771, USA
}

%% `\KeyWords{}' always has to be placed before `\maketitle'.
 %Do NOT move this preamble from here!

\maketitle

\begin{abstract}

Suzaku observed a nearby Seyfert 2 galaxy NGC4945, which hosts 
one of the brightest active galactic nuclei above 20 keV. 
Combining data from the X-ray CCD camera (XIS) and 
the Hard X-ray Detector (HXD), the AGN intrinsic nuclear emission and 
its reprocessed signals were observed simultaneously. 
The intrinsic emission is highly obscured with an absorbing column of 
$\sim 5 \times 10^{24}$~cm$^{-2}$, and was detectable only 
above $\sim 10$~keV. The spectrum below 10~keV is dominated 
by reflection continuum and emission lines from neutral/ionized 
material.
Along with a neutral iron K$\alpha$ line, a neutral iron K$\beta$ and 
a neutral nickel K$\alpha$ line were detected for the first time 
from this source.
The neutral lines and the cold reflection continuum are 
consistent with both originating in the same location. 
The Compton down-scattered shoulder in the neutral Fe-K$\alpha$ line 
is $\sim 10\%$ in flux of the narrow core, 
which confirms that the line originates from reflection
rather than transmission. 
The weakness of the Compton shoulder also indicates
that the reflector is probably seen nearly edge-on.
Flux of the intrinsic emission varied by a factor of $\sim 2$ 
within $\sim 20$~ks, which requires the obscuring material to be 
geometrically thin.
Broadband spectral modeling showed that the solid angle 
of the neutral reflector is less than 
a few~$\times 10^{-2} \times 2\pi$. 
All this evidence regarding the reprocessed signals suggests
that a disk-like absorber/reflector is viewed from a near edge-on angle.

\end{abstract}

%%------------------------------%%
\section{Introduction\label{sec:intro}}
%%------------------------------%%
Our picture of nuclei of Seyfert galaxies relies on 
the unified scheme (\cite{1985ApJ...297..621A}), 
which assumes that the central 
source--- a black hole, an accretion disk, and broad line 
region---is embedded 
within an optically thick molecular torus located 
at a distances larger than $\sim 1$~pc.
Over a certain luminosity range, 
the structure of these main ingredients are considered 
almost identical in all active galactic nuclei (AGNs), 
and the apparent differences among 
individual objects are ascribed to their orientation with respect 
to our line of sight;
the object is classified as a Seyfert 1 
if the line of sight lies within
the opening angle of the torus, and as a Seyfert 2 if observed through
the obscuring material.

Because the nuclear intrinsic light in the optical, ultraviolet, 
and soft X-ray frequencies
is absorbed by the torus, Seyfert 2 AGNs can be seen 
at these energies only through reprocessed emission.
A large population of such obscured AGNs could make 
the Cosmic X-ray background 
(CXB; \cite{1993AdSpR..13..221A}, \cite{1994MNRAS.270L..17M}, 
\cite{1995A&A...296....1C}, \cite{2003MNRAS.339.1095G}, 
\cite{2007A&A...463...79G})
provided that a typical torus is optically thick 
to electron scattering 
and subtends a large solid angle as seen from the central source.
From observations of such obscured AGNs, we can investigate 
the circum-nuclear environment 
(e.g. their geometry, ionization state, etc.), 
thorough the reprocessed emission which would be diluted by the 
intrinsic emission in Seyfert~1's. 
From a fair number of Seyfert~2 AGNs, 
the \textrm{ASCA} satellite actually observed 
evidence for such emission,
e.g. strong iron line complex and hard continua,
in energies from a~few~keV to 10~keV 
(e.g. \cite{1994PASJ...46L.167I}, \cite{1994PASJ...46L..71U}).

Since only hard X-rays can penetrate the obscuring material,
observations in such energy range are crucial 
to the investigation of hidden nuclei, or even to their discovering.
Many of pioneering works on this field were done with 
\textrm{Ginga} satellite, with its hard X-ray sensitivity 
up to $\sim 30$~keV
(e.g. \cite{1989PASJ...41..731K}, \cite{1990Natur.346..544A},
\cite{1993AdSpR..13..221A},\cite{1994ApJ...431L...1U}).
The direct AGN emission, thus observed in hard X-rays, 
were highly absorbed even by 
$N_\mathrm{H}$ of $\gtrsim 10^{24}$~cm$^{-2}$ 
(\cite{1993ApJ...409..155I}), 
where $N_\mathrm{H}$ is equivalent hydrogen column density 
along the line of site.

Subsequently, \textrm{BeppoSAX} and \textrm{RXTE}
observed the hard X-ray emission from a number of Seyfert~2 nuclei
in the $\sim 10$--$100$~keV range
(e.g. \cite{2000MNRAS.316..433I}, \cite{1997A&A...325L..13M}, 
\cite{2003ApJ...589L..17M} and references therein).
In a  fraction of them, the obscuring material was
found to be even opaque to Compton scattering, with 
$N_\mathrm{H} \times \sigma_\mathrm{T}$ reaching a~few
(\cite{2000MNRAS.318..173M}),
where $\sigma_\mathrm{T}$ is the Thomson cross section.
In such ``Compton-thick'' AGNs, 
accurate measurements of the intrinsic emission are quite difficult, 
since it appears only above $\sim 10$~keV 
with a strongly absorbed spectrum. 
In the most extreme cases, the direct AGN emission is 
completely hidden even up to $\sim 100$~keV.

NGC~4945 is a nearby ($\sim$ 3.7~Mpc; \cite{1996A&A...309..705M}),
edge-on (with an inclination 
$i \sim 78^\circ$; \cite{2001A&A...372..463O}) spiral galaxy.
The mass of the central black hole is constrained by H$_2$O megamaser
observations to be $M_{\mathrm{BH}} \sim 1.4 \times 10^6 \MO$
(\cite{1997ApJ...481L..23G}).
The presence of the AGN was first confirmed 
by the \textrm{Ginga} observation 
(\cite{1993ApJ...409..155I}), which revealed 
a rapidly varying nuclear flux 
penetrating through a heavy absorbing column of 
$N_\mathrm{H} \sim 4 \times 10^{24}$~cm$^{-2}$.
Subsequent hard X-ray observations by
the \textrm{OSSE} on-board CGRO (\cite{1996ApJ...463L..63D}),
\textrm{RXTE} (\cite{2000ApJ...535L..87M}),
and \textrm{BeppoSAX} (\cite{2000A&A...356..463G})
showed that NGC~4945 is indeed 
the brightest Seyfert~2 AGN at $\gtrsim 20$~keV.
The nuclear star-burst/super-wind activity was also confirmed 
by infrared and optical observations 
(\cite{1997ApJ...479L..23C}, \cite{1996A&A...308L...1M}).
Observation by \textrm{ASCA} revealed 
a hard continuum below $\sim 10$~keV, and strong iron line complex, 
showing that the reprocessed emission dominates 
in that energy range (\cite{1997ApJ...488..164T}).
Using \textrm{XMM-Newton}, \citet{2002MNRAS.335..241S} found 
both neutral and ionized iron lines 
in the reprocessed nuclear emission, 
but their origins (especially of the latter one) remain rather unclear
even with the superb spatial resolution of  \textrm{Chandra} 
(\cite{2003ApJ...588..763D}).

In the present paper, we report on a 100~ks observation of NGC~4945, 
performed in January 2006 with \textrm{Suzaku} 
(\cite{2007PASJ...59S...1M}).
Thanks to the wide band-pass of \textrm{Suzaku} 
and the good energy resolution at iron K line energy bands, 
we have observed simultaneously the intrinsic 
and the reprocessed emission of NGC~4945, both with a higher accuracy
than in the previous observations.
This helps us to build a consistent picture for the circum-nuclear
re-processing material.

%%------------------------------%%
\section{The \textrm{Suzaku} Observation of NGC~4945}
%%------------------------------%%

Using \textrm{Suzaku}, we observed NGC~4945 over 2006 January 15--17,
for a total duration of 230~ks. 
Both of the instruments on-board, i.e. 
the X-ray Imaging Spectrometer (XIS; \cite{2007PASJ...59S..23K})
and the Hard X-ray Detector (HXD; \cite{2007PASJ...59S..35T}),
were in operation in their normal modes. 
Event files from both instruments were screened using version 1.2 of 
the \textrm{Suzaku} pipeline processing.
We used ``cleaned events'' files in which data of the following
criteria were removed:
data taken with low data rate, 
or with an Earth elevation angle less than $5^{\circ}$, 
or with Earth day-time elevation angles less than $20^\circ$,
or during passages through 
or close to the South Atlantic Anomaly (SAA).
Cut-off rigidity (COR) criteria of $> 6$~GV for the XIS and 
$> 8$~GV for the HXD were also applied.

%%------------------------------%%
  \subsection{XIS Data Analysis}
\begin{figure}
  \begin{center}
    \FigureFile(.9\linewidth,){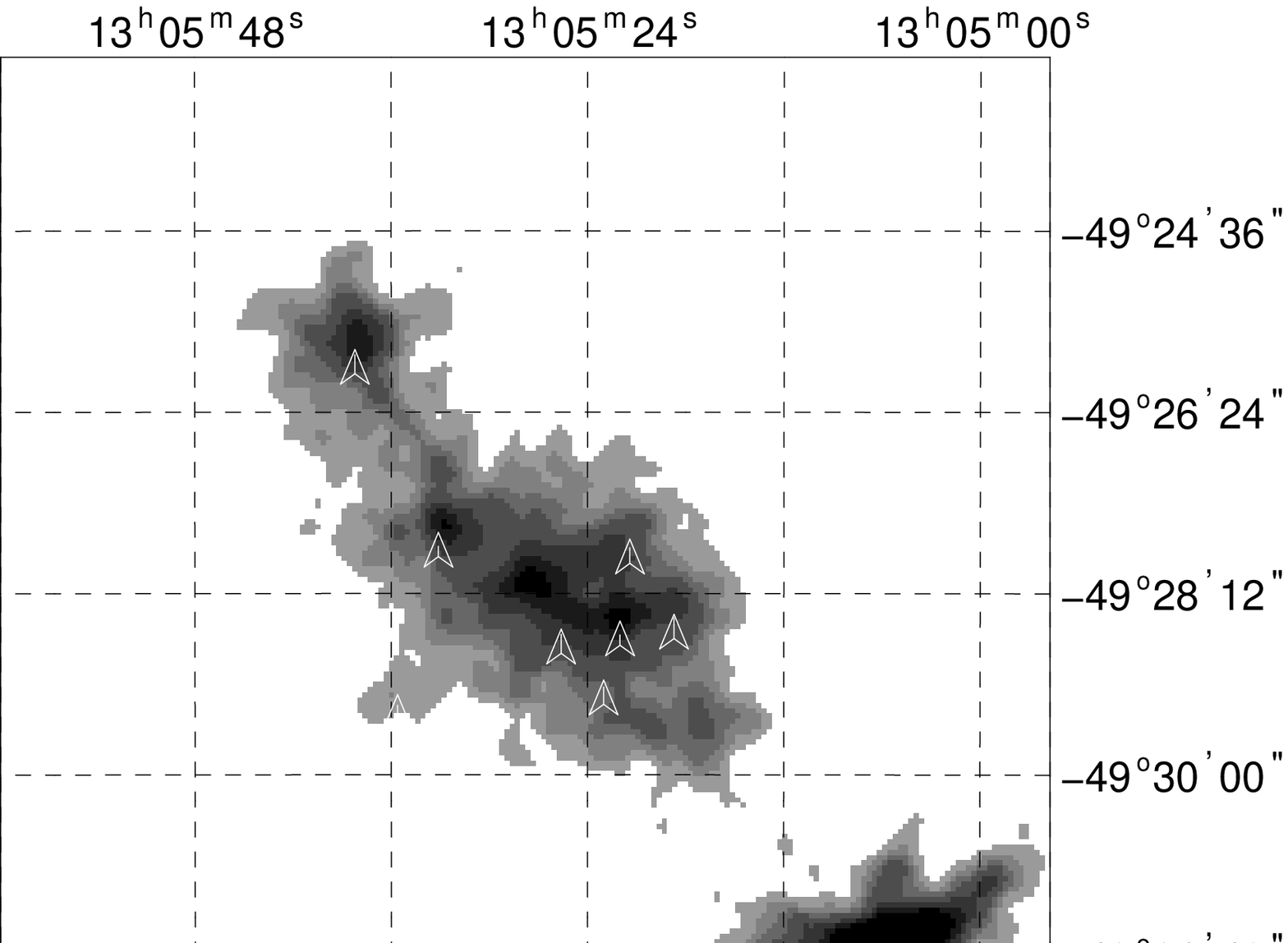}
    \FigureFile(.9\linewidth,){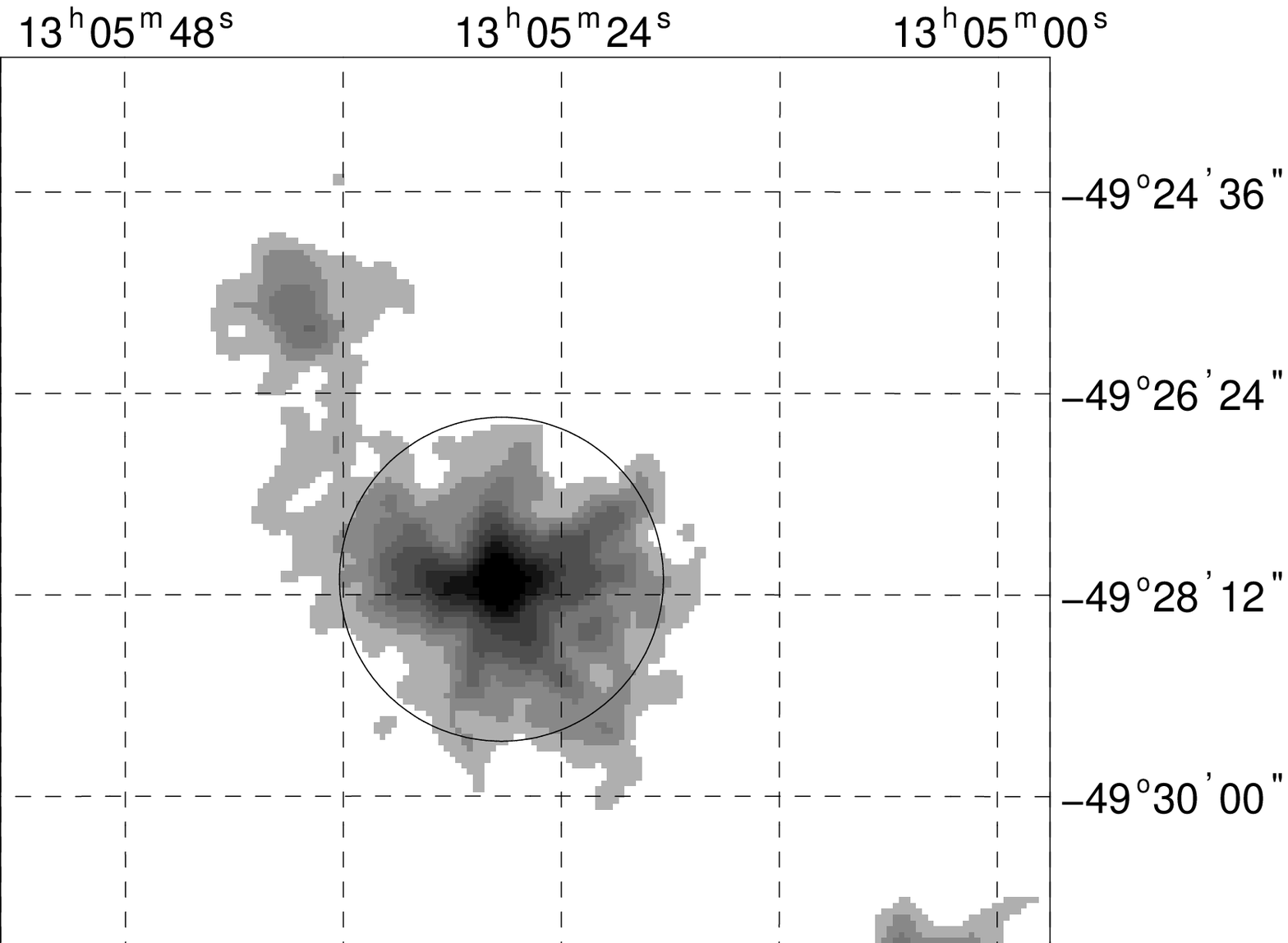}
  \end{center}
  \caption{Background-inclusive \textrm{Suzaku} XIS images 
    of NGC~4945, in the 0.3--4.5~keV (top) 
    and 4.5-9.0~keV band (bottom).
    Only events from XIS0 sensor are used.
    White arrows in the top panel indicate individual point sources
    detected by \textrm{XMM-Newton} (\cite{2002MNRAS.335..241S}).
    Black circle in the bottom panel indicates the source 
    extraction region. Its radius of $1'.45$ corresponds to 2.00~mm
    on the sensor, or 1.56~kpc in physical size at the distance of 
    3.7~Mpc.
  }\label{fig:xisimage}
\end{figure}

%%------------------------------%%
In analyzing the XIS data, 
we used only events with grades 0, 2, 3, 4 and 6. Hot and 
flickering pixels were removed using \texttt{cleansis} software.
From each XIS sensor, we omitted those periods which suffered from 
telemetry saturation.
Figure \ref{fig:xisimage} shows the XIS image of NGC~4945, 
in energy ranges of 0.3--4.5~keV (top) and 4.5--9.0~keV (bottom).
While several sources can be seen in the lower energy band, 
the AGN emission is quite dominant in the higher energy band. 
In the same region, \citet{2002MNRAS.335..241S} 
detected non-nuclear point sources 
which are indicated with white arrows 
in figure \ref{fig:xisimage}~(top).
We analyzed the \textrm{XMM-Newton} data of these sources, 
to find that their contamination to the \textrm{Suzaku} AGN spectra 
are negligible in energies higher than $\sim 4.5$~keV.

The bright source in the SW region 
in figure \ref{fig:xisimage} has not been
observed by any previous observations. 
According to its luminosity, this transient source could be 
a ultra luminous X-ray source (ULX) associated with NGC~4945. 
Readers are referred to \citet{Isobe2007} for 
detailed analysis of this source.

AGN spectra were extracted from the four XIS sensors 
within a circular region of 
$1'.45$ radius (corresponding to 1.56~kpc in physical size 
at a distance of 3.7~Mpc), 
indicated by a black circle in figure \ref{fig:xisimage} (bottom). 
The background spectra were
extracted from source free regions of the same CCDs.
We made the response matrices and ancillary response 
matrices using \texttt{xisrmfgen} and \texttt{xissimarfgen} 
(\cite{2007PASJ...59S.113I}) , respectively. 
In the following spectral analysis, we add the spectra and 
responses of the front-illuminated (FI) chips (XIS 0, 2, 3), 
and use data over an energy range of 4.5--9.0~keV 
to avoid possible contamination by other sources. 
We use an energy range of 4.5--8.0~keV for the back-illuminated (BI) 
XIS1 chip, due to its high background above this energy.
A total net exposure of 98.8~ks was obtained from each XIS sensor, 
with source count rates of 
$(8.83 \pm 0.1) \times 10^{-2}$~cts~s$^{-1}$ for the FI chips 
and $(1.88 \pm 0.05) \times 10^{-2}$~cts~s$^{-1}$ for the BI chip. 
The source spectra were re-binned so that each bin includes 
at least 30 counts to enable usage of the  $\chi^2$  fit statistics.

%%------------------------------%%
\subsection{HXD Data Analysis}
%%------------------------------%%

%%------------------------------%%
%\subsubsection{HXD-PIN}
%%------------------------------%%
The HXD-PIN spectrum was extracted from the ``cleaned events'' 
file described above.
In order to estimate  non X-ray background (NXB) events, 
we used a time-dependent instrumental background event file 
provided by the instrumental team. 
We extracted the source and the background spectra
after applying an identical set of good time intervals  (GTIs) to 
both event files. 
The source spectrum was further corrected for detector dead time
using ``pseudo events'' (\cite{2007PASJ...59S..53K}).
The dead time was  $\lesssim 5\%$, and the achieved 
total net exposure was 69.5~ks.

The contribution of cosmic X-ray background 
(CXB; \cite{1987PhR...146..215B}) 
was estimated using the HXD-PIN response for diffuse emission.
We adopted the form of CXB determined by \textrm{HEAO~1}
(\cite{1999ApJ...520..124G}); 
$9.0 \times 10^{-9} \times (E/3~\mathrm{keV})^{-0.29} \times 
\exp{(-E/40~\mathrm{keV})}$~erg~cm$^{-2}$~s$^{-1}$~str$^{-1}$~keV$^{-1}$,
where $E$ is the photon energy.
The CXB count rate is estimated to be $\sim 9\%$
of the net source signals.
Recent works find that the CXB normalization 
is higher by $\sim 10$--15\% at $\sim 10$--100~keV, 
than that of \citet{1999ApJ...520..124G} 
(\cite{2003A&A...411..329R}, \cite{2007A&A...467..529C}).
This level of uncertainty, however, corresponds to only $\sim 1\%$
of the net source flux of NGC~4945, 
and does not affect the subsequent results on the 
high energy spectrum.

\begin{figure}
  \begin{center}
    \FigureFile(.9\linewidth,){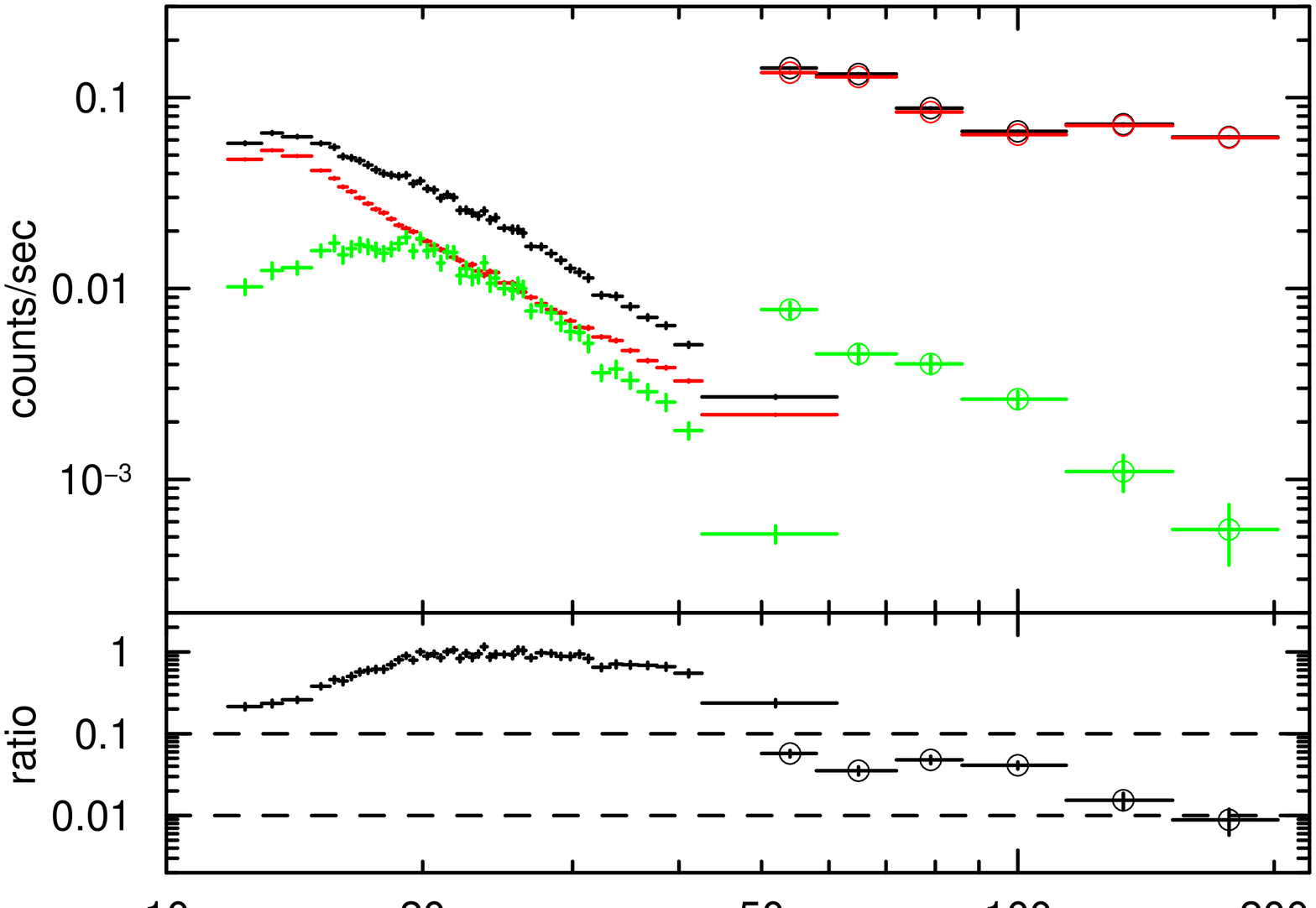}
  \end{center}
 \caption{HXD spectra from the present Suzaku observation of NGC~4945.
    The PIN and GSO data points are indicated 
    by crosses and circles, respectively.
    In the upper panel, the total spectra 
    (source plus background) are shown in black, 
    the estimated background spectra in red,  
    and the net source (total minus background) spectra in green.
    The lower panel shows the ratio of the net source to the background
    spectra.
  }\label{fig:hxd_srcbgd_comp}
\end{figure}

We show in figure \ref{fig:hxd_srcbgd_comp} the energy spectra of 
NGC~4945 observed with HXD-PIN (below 60~keV), in comparison with 
HXD-PIN background which includes both NXB and the CXB.
As shown in the lower panel of figure \ref{fig:hxd_srcbgd_comp}, 
the net source counts exceed $\sim 20$\% of the background 
over the entire HXD-PIN energy range. 
In the following spectral analysis, 
we use an energy range of 12--60~keV, in which the source signals are 
detected at $> 8 \sigma$ above the background in each of 
appropriately summed energy bins in the spectrum.
We use the response matrix of version \texttt{2006-8-14},
which is appropriate for the HXD nominal aiming position. 
The background-subtracted source count rate in the energy range,
with statistical error, is $0.299 \pm 0.003$~cts~s$^{-1}$, 
which is $\sim 65$\% of the 
background rate (NXB plus CXB, $0.461 \pm 0.001$~cts~s$^{-1}$).

%%------------------------------%%
%\subsubsection{HXD-GSO}
%%------------------------------%%

The source and background spectra from HXD-GSO 
were extracted in the same way as for HXD-PIN. 
We further applied the instrumental dead-time correction to 
the background spectrum as well, which was not necessary 
in the case of HXD-PIN.
We present in figure \ref{fig:hxd_srcbgd_comp} 
the AGN energy spectrum taken with HXD-GSO (above 50~keV).
The HXD-GSO background does not include the CXB, 
which is negligible in this case.
Figure \ref{fig:hxd_srcbgd_comp} (bottom) show that 
even in the HXD-GSO range, the signal exceeds 2\% of the 
background up to $\sim 150$~keV, 
although it decreases to $\lesssim 1$\% at $\sim 200$~keV.
Systematic studies on the modeled background spectrum of HXD-GSO
shows that its uncertainty is $\sim 2\%$ (\cite{Takahashi2007}),
so that we can claim the detection of this source 
up to 150~keV at present.
We use the response matrix of version \texttt{2006-3-21} 
and arf file \texttt{ae\_hxd\_gsohxnom\_crab\_20070502.arf} 
(\cite{Takahashi2007}), 
which are appropriate for the HXD nominal aiming position. 
In the subsequent spectral analysis, we utilize the GSO data only 
in the 70--120~keV range, because of the current uncertainties in the
HXD-GSO background and response, in the $> 120$~keV and 50--70~keV 
energies, respectively. 

\subsection{Broadband Spectrum}

\begin{figure}
  \begin{center}
    \FigureFile(.9\linewidth,){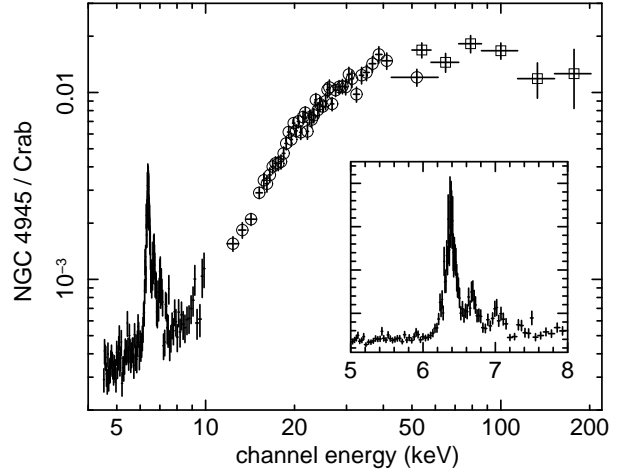}
  \end{center}
  \caption{The background-subtracted Suzaku spectra of the nucleus 
    of NGC~4945, normalized to those of the Crab nebula. 
    The XIS-FI, PIN and GSO data points are marked with crosses, 
    circles and squares, respectively. 
    (Inserted figure) A zoom of the 5--8~keV band, 
    with the flux in linear scale.
    The XIS-BI data are omitted just for clarity.
 }\label{fig:crabratio}
\end{figure}

Figure \ref{fig:crabratio} shows the source spectra 
in the 4.5--200~keV band, normalized to those of the Crab nebula
taken with Suzaku on 2006 September 15 
at the HXD nominal aiming position.
The total net exposure onto the Crab was $\sim 12$~ks with the XIS, 
and $\sim 10$~ks with the HXD.
Since the energy spectrum of the Crab is known to be
a single power-law with a photon index of $\sim 2.1$, 
the ratio spectrum can be regarded as an approximate 
$\nu F \nu$ spectrum.
Figure \ref{fig:crabratio} reveals an extremely hard spectrum 
in the $\sim 10$--30~keV band, which indicates 
that the underlying emission is attenuated by a considerable 
degree of absorbing column. 
Above $\sim 40$~keV, the Crab ratio stays rather constant up to 
$\sim 200$~keV, without showing clear evidence for any 
high-energy spectral cutoff.

The continuum below $\sim 10$~keV is softer 
but still quite hard ($\Gamma \sim 1$), 
so that it can be better interpreted as a reflected continuum
rather than an intrinsic emission.
%In addition, strong emission line features are clearly seen 
%around 6--8~keV, in the inserted figure.
This inference is supported also by the strong emission 
line features, which are clearly seen around 6--8~keV 
in the inserted figure.

%--------------------------------%
\section{Broad-Band Spectral Fitting \label{sec:spec_fit}}
%--------------------------------%

%--------------------------------%
\subsection{Baseline Continuum Model}
%--------------------------------%
As described in $\S$\ref{sec:intro}, the nuclear intrinsic 
emission is considered to appear only above $\sim 10$~keV, 
while its re-processed emission dominates the energy 
spectrum below 10~keV.
As an approximation to the 4.5--50~keV spectrum, we 
then adopted an baseline continuum (BC) model 
consisting of an absorbed power-law 
with high energy exponential cutoff ($\gtrsim 100$~keV), 
and a reflection continuum by a cold matter which is absorbed 
by the Galactic column 
($1.57 \times 10^{21}$~cm$^{-2}$; \cite{1979AuJPA..47....1H}) only.
We multiplied the first component by an energy-dependent 
factor of the form $\exp{\{-N_{\mathrm{H}} \sigma(E)\}}$, 
where the cross-section $\sigma(E)$ includes the effects of both 
photo-electric absorption (model \texttt{wabs} in XSPEC)
and Compton scattering (model \texttt{cabs} in XSPEC).
The latter component is based on an assumption that some fraction, 
$f_\mathrm{refl}$, of the primary power-law is reflected 
from cold material, and reaches us without further absorbed
by the obscuring matter which affects the first component.
We modeled the cold reflection using \texttt{pexrav} in XSPEC 
(\cite{1995MNRAS.273..837M}), where the continuum shape $\Gamma$
was tied to that of the intrinsic power-law. 
The inclination ($i$) of the reflector 
was set to 60$^\circ$, and the iron abundance 
($A_{\mathrm{Fe}}$) was fixed at 1. 
%The relative normalization between the reflected and the 
%transmitted component, $R$, is set to -1, 
%which forces the model to produce only the reflected component.
The \texttt{pexrav} model was used in such a way that
it produces the reflected photons only, 
because the illuminating primary continuum is separately modeled.
The model reproduce a neutral iron edge structure at $\sim 7.1$~keV, 
while it does not include any fluorescent lines, 
although their occurrence is physically expected.

\begin{figure}
  \begin{center}
    \FigureFile(.9\linewidth,){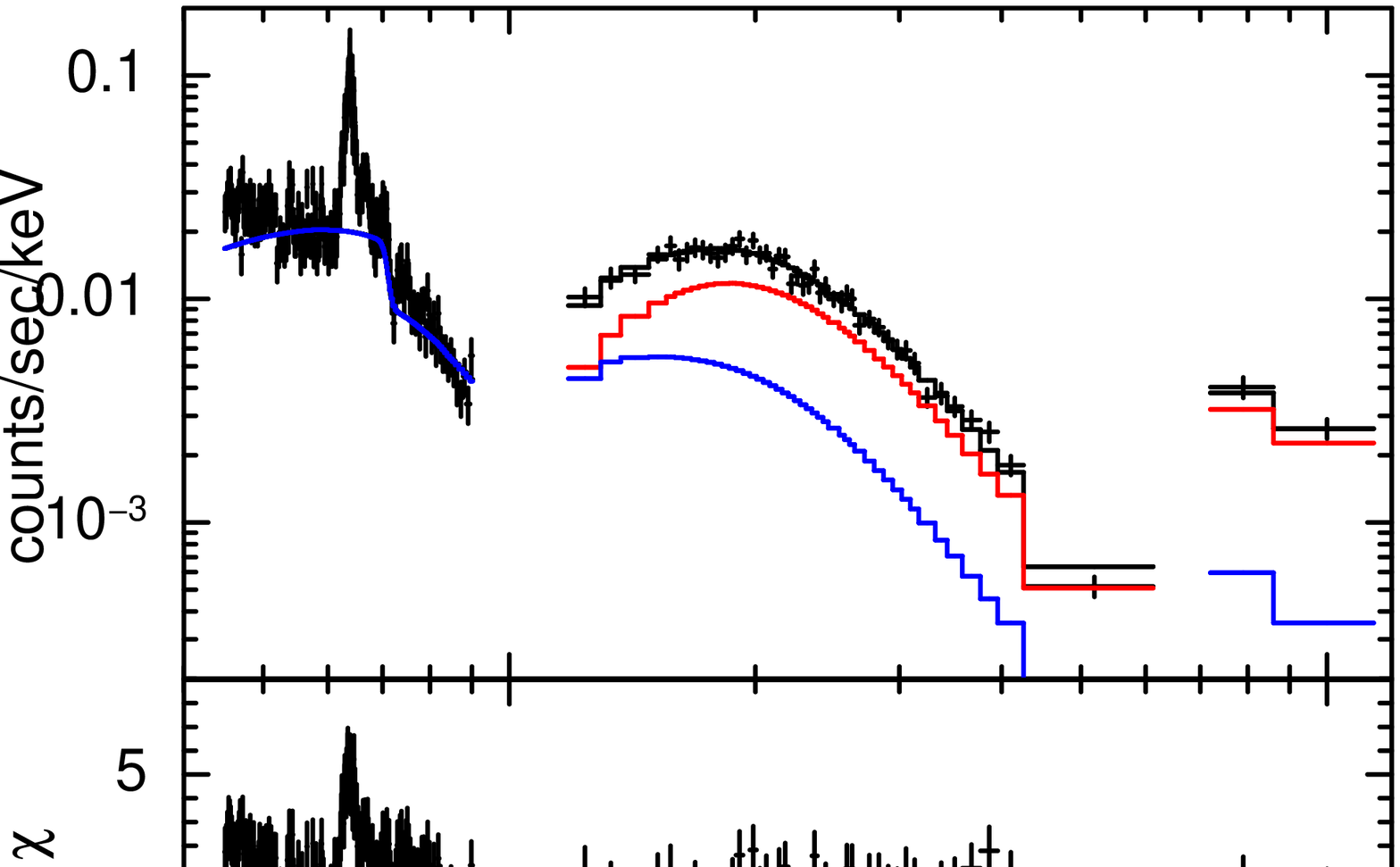}
    \FigureFile(.9\linewidth,){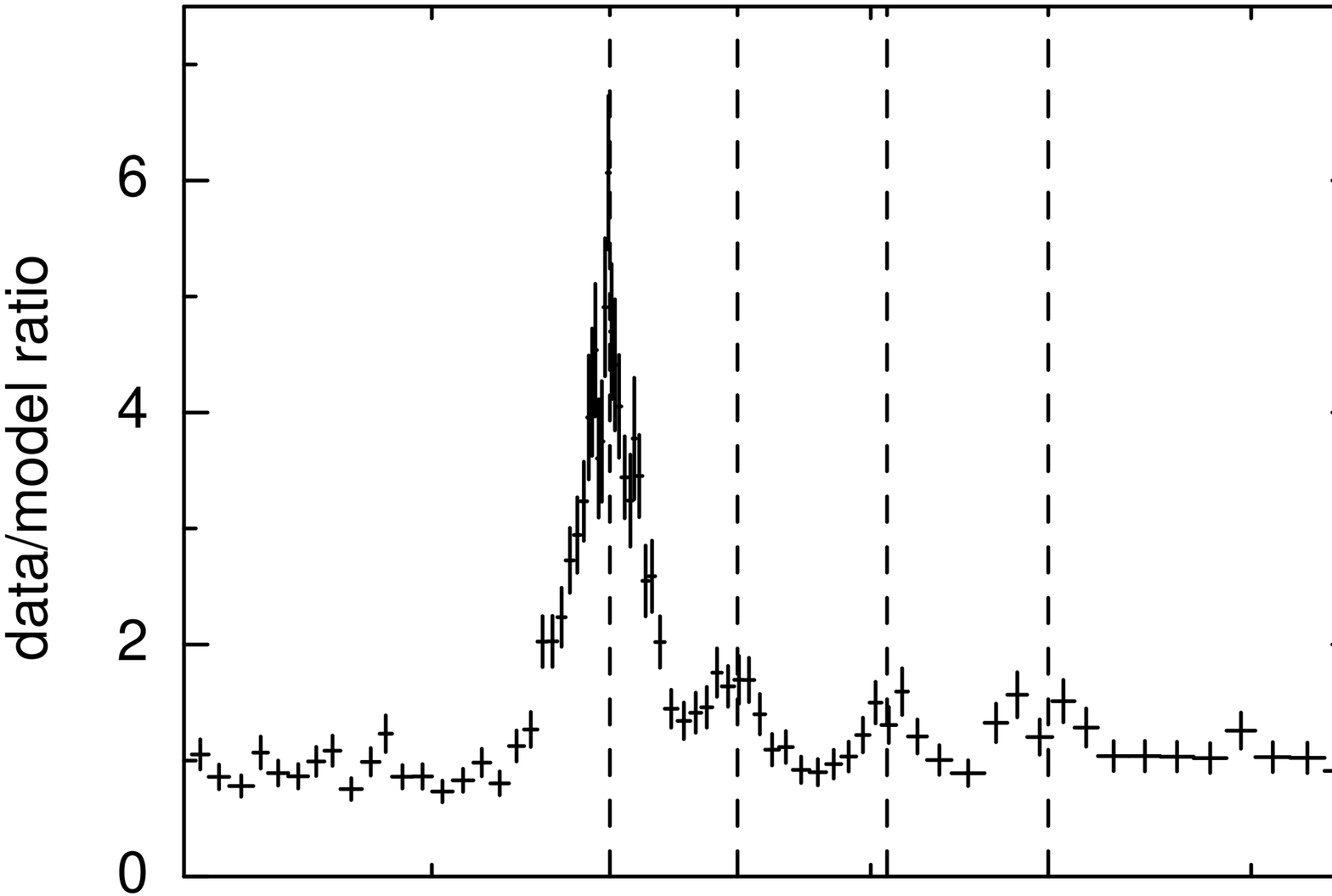}
  \end{center}
  \caption{(Top) The broad band (4.5--120 keV) 
    Suzaku spectrum of NGC~4945. The data and the baseline
    continuum model (see text) are shown. 
    The models are shown in colored lines;
    the absorbed intrinsic power-law in red, 
    the reflection component in blue, 
    and their sum in black.
    (Bottom) The data to model ratios
    from this fit in the 5.5--8.2~keV band.
  }\label{fig:xis_pin_spec}
\end{figure}

The data and the best-fit BC model are shown in 
the top panel of figure \ref{fig:xis_pin_spec}.
The spectrum up to 10~keV is dominated by the cold reflection 
component, while the luminous intrinsic power-law appears 
above that energy.
%The we exclude the high energy cutoff, 
The cutoff energy is constrained to be $200^{+150}_{-50}$~keV 
in the fit. However, when the GSO background systematic error
($\sim 2\%$) are taking into account, we get only its lower limit
of $\sim 80$~keV.

%The ratio reveals several emission lines, 
%at $\sim 6.40$~keV, $\sim 6.70$~keV, $\sim 7.05$~keV, 
%and possibly at $\sim 7.47$~keV. 
The bottom panerl of figure \ref{fig:xis_pin_spec} shows the 
data/model ratio around the iron line region.
It reveals several emission lines, at $\sim 6.40$~keV, $\sim 6.70$~keV, 
$\sim 7.05$~keV, and possibly at $\sim 7.47$~keV. 
We therefore added Gaussians to represent those four features,
and found that all of them are statistically significant 
($\gtrsim 3\sigma$ even with the $\sim 7.47$~keV one).
In addition, the fitted center energies of three of them,
$6.396^{+0.005}_{-0.005}$, $7.06^{+0.03}_{-0.04}$, 
and $7.45^{+0.05}_{-0.07}$~keV
(here and hereafter in the source rest-frame energies, 
assuming $z = 0.0019$),
agree with those of 
Fe~I~K$\alpha$, Fe~I~K$\beta$ and Ni~I~K$\alpha$ lines, respectively.
This indicate that they originate as fluorescence 
from cold materials.
Note that the latter two emission lines have not been detected 
by any previous missions.
Apart from the three neutral lines, 
the emission line at $\sim 6.676^{+0.025}_{-0.020}$~keV agrees
in energy with Fe~XXV~K$\alpha$ line, 
suggesting the presence of another re-processing material 
which is highly-ionized.

\begin{figure}
 \begin{center}
  \FigureFile(.9\linewidth,){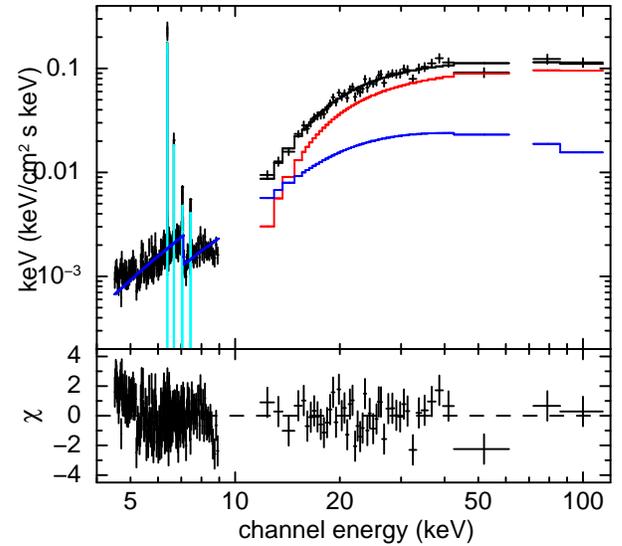}
 \end{center}
 \caption{(Top) The broad band (4.5--120 keV) $\nu F \nu$ 
       spectrum of NGC~4945, on which BC 
       model is superposed. The data are shown in black 
       points and the models in colored lines; 
       the absorbed intrinsic power-law in red, 
       the cold reflection component in blue,
       and emission lines in cyan. 
       (Bottom) The residuals from this fit.
 }\label{fig:ic_eeuf_dchi}
\end{figure}

We show in figure \ref{fig:ic_eeuf_dchi}
the $\nu F \nu$ spectrum,  on which the fitted model is superposed.
The best-fit parameters of the BC plus Gaussian lines 
are summarized in table \ref{tab:first}. 
The model reproduces the observed continuum marginally 
successfully ($\chi^2/$d.o.f. $= 330/242 \sim 1.4$), 
though systematic residuals are recognized 
(figure \ref{fig:ic_eeuf_dchi} bottom) around $\sim 5$~keV
and $\sim 9$~keV.

\subsection{BC plus Thermal Plasma Model\label{sec:ictherm}}

\begin{figure}
 \begin{center}
  \FigureFile(.9\linewidth,){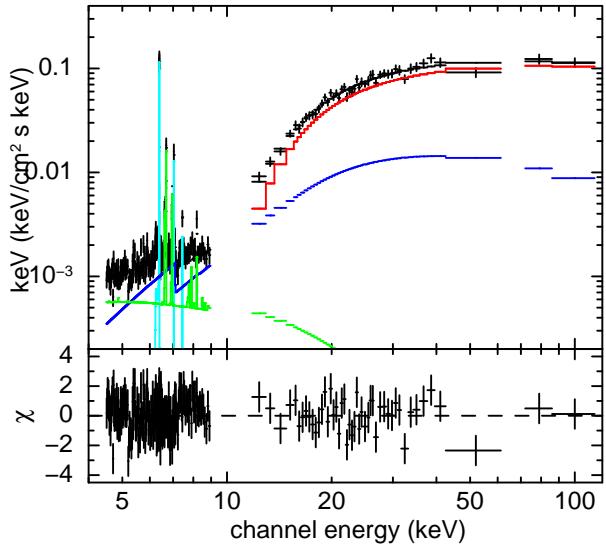}
 \end{center}
 \caption{The same as figure \ref{fig:ic_eeuf_dchi},
       but for the BC + thermal plasma model.
       The data are shown in black 
       points and the models in colored lines in the top panel; 
       the absorbed intrinsic power-law in red, 
       the cold reflection component in blue,
       the thermal plasma in green, and emission lines in cyan. 
 }\label{fig:ic_therm_eeuf_dchi}
\end{figure}

The ionized material from which the Fe~XXV~K$\alpha$ line originates 
may alternatively be a hot plasma collisionally heated 
by the star-burst activity around the nucleus 
(\cite{2002MNRAS.335..241S}).
We thus added to the BC model a hot plasma component
(\texttt{mekal} in XSPEC). 
We separately modeled the three neutral lines;
the Fe~I~K$\beta$ and Ni~I~K$\alpha$ lines by single Gaussians,
while we used the following model for the strong Fe~I~K$\alpha$ line.
The model consists of two Gaussian lines,  
K${\alpha}_1$ (6.404~keV) and K${\alpha}_2$ (6.391~keV), 
and also includes a Compton down-scattered shoulder,
which is physically expected when the line is produced in 
optically-thick re-processing materials (\cite{2002MNRAS.337..147M}).
The profile of the shoulder was calculated according to 
\citet{1979ApJ...228..279I}, 
while its normalization was allowed to vary. 
The ratio between the normalization of the shoulder and that of the 
narrow core is quoted as $f_\mathrm{CS}$ hereafter.
The center energies of the Fe~I~K$\beta$ and Ni~I~K$\alpha$ line 
were fixed at 7.06~keV and 7.47~keV, 
and their normalization were also fixed at 12\% and 5\% 
of that of Fe~I~K$\alpha$, respectively.
The fit goodness was significantly improved ($\chi^2$/d.o.f. = 257/247)
compared to the BC model,
mainly due to the inclusion of the hot plasma.
We show in figure \ref{fig:ic_therm_eeuf_dchi}
the $\nu F \nu$ spectrum on which the fitted model is superposed.
The best-fit parameters are summarized in table \ref{tab:first}.

The Fe~I~K$\alpha$ line has an equivalent width of 
$1.3^{+0.5}_{-0.2}$~keV with respect to the cold reflection component.
This is consistent with the prediction in the case when
the line and reflected continuum originate at the same location, 
possibly a visible part of the putative torus 
(e.g. \cite{1996MNRAS.280..823M}).
With the column density of 
$N_\mathrm{H} \gtrsim \mathrm{a~few} \times 10^{24}$~cm$^{-2}$, 
the electron scattering suppresses the intrinsic emission 
by a factor of $\exp\{- N_\mathrm{H} \cdot {\sigma}_\mathrm{T} \}$.
Thus $f_\mathrm{refl}$, calculated as a ratio between 
the normalization of \texttt{pexrav} and that of the absorbed 
power-law, is in reality much smaller than would be read on figure 
\ref{fig:ic_therm_eeuf_dchi}, 
and becomes $\sim \mathrm{a~few}~\times 10^{-3}$. 
The Compton shoulder parameter $f_\mathrm{CS}$ is constrained 
to be $\sim 0.11^{+0.07}_{-0.10}$, 
which is consistent with those from re-processing by 
a near-Compton-thick reflector
though slightly smaller than expected value ($\sim 0.20$); 
see \citet{2002MNRAS.337..147M} for calculation, 
and e.g. \citet{2006MNRAS.368..707P} and 
\citet{2005MNRAS.360.1123P} for observational evidence 
in NGC~1068 and MKN~3, respectively.

\subsection{BC plus Scattered Continuum Model}

Recent studies with high resolution instruments show radiative
recombination continua, which provide clear evidence for photo-ionized gas.  
These atomic features are typically seen in the highest 
signal--to--noise spectra of obscured AGNs
(e.g. the compilation of \cite{2006A&A...448..499B}, 
\cite{2001ApJ...556....6Y}; \cite{2002ApJ...575..732K}). 
According to detailed calculations of the column, ionisation state 
and covering fraction of the gas required to produce the radiative
recombination continuua and the copious low energy photo-ionised 
lines in Seyfert 2's, such material must necessarily also 
scatter some fraction $f_\mathrm{scatt}$ of the photo-ionizing flux, 
producing a continuum which has approximately the same shape as 
the incident AGN emission, i.e. a power law (\cite{2006A&A...448..499B}). 
This continuum can be seen directly in regions of the spectrum away 
from the strong line emission (e.g. Mkn 3: \cite{2005MNRAS.360..380B}). 
Hence we explore here the possibility that the Fe XXV K$\alpha$ 
line seen in NGC~4945 is from photo-ionized rather than 
collisionally ionized material, 
although the RGS data, hampered by poor signal--to--noise ratio of 
this object at low energies, did not give strong evidence for 
such a photo-ionized gas (\cite{2007MNRAS.374.1290G}).
%though the limited 
%signal--to--noise in this object at low energies mean that 
%there is no unambiguous requirement for photo-ionized gas from 
%the RGS data (\cite{2007MNRAS.374.1290G}).

\begin{figure}
  \begin{center}
    \FigureFile(.9\linewidth,){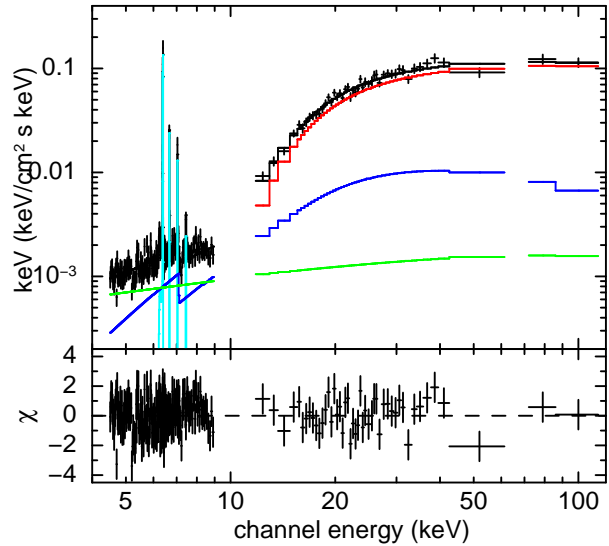}
  \end{center}
  \caption{The same as figure \ref{fig:ic_eeuf_dchi},
       but for the BC + scattered power-law model.
       The green line shows the continuum scattered by 
       a highly ionized scatterer.
  }\label{fig:ic_scpl_eeuf_dchi}
\end{figure}

\begin{figure}
  \begin{center}
    \FigureFile(.9\linewidth,){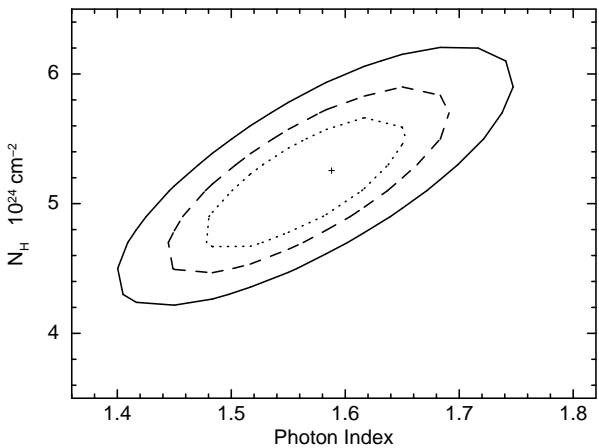}
  \end{center}
  \caption{Confidence (68\%, 90\%, 99\%) contours of the photon index
    $\Gamma$ of the primary continuum vs. the absorption column 
    $N_\mathrm{H}$,
    calculated under the ``BC + Scattered Continuum'' model.
  }\label{fig:cont_gam_nh}
\end{figure}

We thus added to the BC model, 
another power-law absorbed only by the Galactic column, 
with its photon index set to be the same as the primary one.
We modeled the neutral lines in the same way as in \S\ref{sec:ictherm},
while the Fe~XXV~K$\alpha$ line with a single Gaussian, 
allowing its energy to vary. 
This model has also reproduced the data successfully,
with a significantly better $\chi^2$ than the previous one 
($\chi^2$/d.o.f. = 241/246). 
We show in figure \ref{fig:ic_scpl_eeuf_dchi}
the $\nu F \nu$ spectrum on which the fitted model 
is superposed.
The obtained best-fit parameters are shown in table~\ref{tab:first}.
We show in figure \ref{fig:cont_gam_nh} confidence contours of the
primary continuum slope, $\Gamma$, vs the absorption column density,
calculated under this model.

The equivalent width of the Fe~I~K$\alpha$ line, 
with respect to the cold reflection, 
is calculated at 1.6$^{+0.3}_{-0.3}$~keV.
Adding another Gaussian at 6.96~keV, corresponding to 
Fe~XXVI~K$\alpha$, did not improve the fit.
The equivalent widths of the Fe~XXV~K$\alpha$ 
and Fe~XXVI~K$\alpha$ lines,
with respect to the scattered power-law, 
is $\sim 0.5$~keV and $< 0.1$~keV, respectively.

In either of the BC~+~thermal or the BC~+~scattered~PL modeling, 
an unphysical trade-off between cold reflection continuum 
and thermal/scattered~PL component may happen. 
We thus calculated the 5--10~keV flux 
of the cold reflection continuum, and obtained 
7.8$^{+1.0}_{-1.2} \times 10^{-13}$ erg~s$^{-1}$~cm$^{-2}$, 
for each of the BC~+~thermal or BC~+~scattered~PL model.
This value agrees well with that obtained
using \textrm{Chandra} ACIS, 
from the nucleus and a surrounding region of $\sim 5''$ scale 
(8.0$^{+1.3}_{-1.5} \times 10^{-13}$ erg~s$^{-1}$~cm$^{-2}$, 
\cite{2003ApJ...588..763D}).

\begin{table*}
  \caption{Best-fit parameters of the three models (see text)
           to the Suzaku XIS, HXD-PIN, and HXD-GSO spectrum of NGC~4945.
}
 \label{tab:first}
 \begin{center}
  \begin{tabular}{llrrr}
   \hline\hline
   Model Component & Parameter & BC + lines & BC + Thermal & BC + Scattering\\
   \hline
   Intrinsic PL     & $\Gamma$%
                       & $1.6^{+0.1}_{-0.1}$  
                       & $1.5^{+0.2}_{-0.1}$   
                       & $1.6^{+0.1}_{-0.2}$ \\
                    & $N_\mathrm{PL}$\footnotemark[*]%
                       & $3.0^{+1.5}_{-1.5}$   
                       & $1.4^{+0.7}_{-0.7}$    
                       & $1.7^{+1.0}_{-0.7}$  \\

                    & $E_\mathrm{cut}$\footnotemark[$\dagger$]%
                       & $200^{+150}_{-50}$\footnotemark[$\ddagger\ddagger$]
                       & $150^{+100}_{-50}$\footnotemark[$\ddagger\ddagger$]
                       & $190^{+150}_{-50}$\footnotemark[$\ddagger\ddagger$] \\
                    & $N_\mathrm{H}$\footnotemark[$\ddagger$]%
                       & $6.0^{+0.5}_{-0.5}$   
                       & $5.4^{+0.4}_{-0.4}$    
                       & $5.3^{+0.4}_{-0.9}$      \\[1.5ex]
   Reflection       & $f_\mathrm{refl}$%
                       & $3^{+3}_{-2} \times10^{-3}$ 
                       & $3^{+2}_{-1} \times10^{-3}$ 
                       & $ 3^{+1}_{-1} \times10^{-3}$  \\[1.5ex]
   Mekal            & k$T$\footnotemark[\S]%
                       & -                 
                       & $7.1^{+1.0}_{-0.9}$   
                       & -                   \\
                    & norm\footnotemark[$\|$]%
                       & -                 
                       & $1.2^{+0.1}_{-0.2} \times 10^{-3}$ 
                       & -          \\[1.5ex]
   Scattered PL     & $f_{\mathrm{scat}}$%
                       & -                 
                       & -                  
                       & $\sim 2 \times10^{-4}$  \\[1.5ex]
   Fe~I~K$\alpha$   & $E_\mathrm{c}$\footnotemark[\#]%
                       & $6.396^{+0.005}_{-0.005}$ 
                       & see text  
                       & see text            \\
                    & $I$\footnotemark[**]%
                       & $29^{+2}_{-2}$    
                       & $31^{+2}_{-2}$     
                       & $32^{+2}_{-2}$      \\
                    & EW\footnotemark[$\dagger\dagger$]%
                      & $\sim 0.6$    
                      & $1.3^{+0.5}_{-0.2}$\footnotemark[\S\S] 
                      & $1.6^{+0.3}_{-0.3}$\footnotemark[\S\S]\\[1.5ex]
   CS               & $f_\mathrm{CS}$%
                       & -                 
                       & $0.11^{+0.07}_{-0.10}$   
                       & $0.10^{+0.08}_{-0.09}$    \\[1.5ex]
   Fe~XXV~K$\alpha$ & $E_\mathrm{c}$\footnotemark[\#]%
                       & $6.675^{+0.025}_{-0.020}$ 
                       &  -                 
                       & $6.683^{+0.015}_{-0.015}$  \\
                    & $I$\footnotemark[**]%
                       & $6.0^{+1.5}_{-1.2}$  
                       &  -                 
                       & $8.4^{+1.0}_{-1.5}$     \\
                    & EW\footnotemark[$\dagger\dagger$]%
                       & $\sim 0.2$        
                       &  -                 
                       & $0.5^{+*}_{-*}$    \\[1.5ex]
   Fe~I~K$\beta$    & $E_\mathrm{c}$\footnotemark[\#]%
                        & $7.06^{+0.03}_{-0.04}$ 
                        & 7.06 fixed      
                        & 7.06 fixed          \\
                    & $I$\footnotemark[**]%
                        & $2.9^{+1.2}_{-1.3}$  
                        &   -\footnotemark[\#\#]       
                        & -\footnotemark[\#\#]           \\
                    & EW\footnotemark[$\dagger\dagger$]%
                        & $\sim 0.1$   
                        & $\sim 0.2$\footnotemark[\S\S]             
                        & $\sim 0.2$\footnotemark[\S\S]   \\[1.5ex]
   Ni~I~K$\alpha$   & $E_\mathrm{c}$\footnotemark[\#]%
                        & $7.45^{+0.05}_{-0.07}$ 
                        & 7.47 fixed         
                        & 7.47 fixed          \\ 
                    & $I$\footnotemark[**]%
                        & $2.7^{+1.0}_{-1.0}$  
                        & -\footnotemark[***]               
                        & -\footnotemark[***]          \\
                    & EW\footnotemark[$\dagger\dagger$]%
                        & $\sim 0.1$   
                        & $\sim 0.1$\footnotemark[\S\S]         
                        & $\sim 0.1$\footnotemark[\S\S]  \\[1.5ex]
   $\chi^{2}$/d.o.f. &   & 330/242   & 257/247   & 241/246    \\
      \hline\hline
\multicolumn{5}{@{}l@{}}{\hbox to 0pt{\parbox{180mm}{\footnotesize
\par\noindent
       \footnotemark[*]
                $\mathrm{photons~keV^{-1}~cm^{-2}~s^{-1}}$~at 1 keV.
       \footnotemark[$\dagger$]keV.
       \footnotemark[$\ddagger$]$10^{24}$ cm$^{-2}$.
       \footnotemark[\S]keV.
       \footnotemark[$\|$]
       \footnotemark[\#]Center energy in keV.
\par\noindent
       \footnotemark[**]$10^{-6}$~photons~cm$^{-2}$~s$^{-1}$.  
       \footnotemark[$\dagger\dagger$]keV.
       \footnotemark[$\ddagger\ddagger$]
                 Only lower limit of $\sim 80$~keV is obtained
                 when the GSO \\ background systematics 
                 are taking into account.
\par\noindent
       \footnotemark[\S\S]
                 EW with respect to the cold reflection continuum. 
       \footnotemark[$\|\|$]Upper limit.
\par\noindent
       \footnotemark[\#\#]Set equal to 12\% of that of Fe~I~K$\alpha$.
       \footnotemark[***]Set equal to 5\% of that of Fe~I~K$\alpha$.
     }\hss}}
  \end{tabular}
 \end{center}
\end{table*}

%--------------------------------%
%\subsection{Broad-Band Spectrum}
%--------------------------------%
%--------------------------------%
%\subsection{The iron line profiles}
%--------------------------------%

%--------------------------------%
\section{Broadband Time Variability\label{sec:timing}}
%--------------------------------%

\begin{figure}
  \begin{center}
    \FigureFile(.9\linewidth,){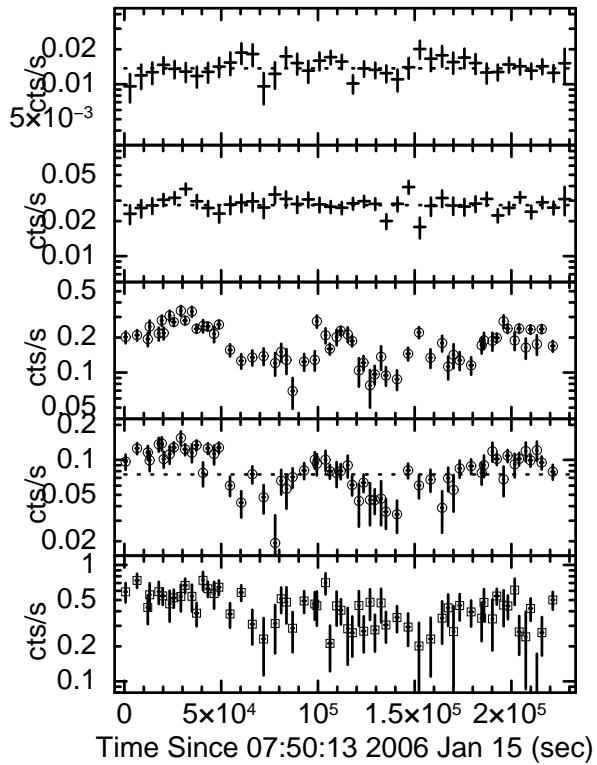}
  \end{center}
  \caption{Background-subtracted light curves of NGC~4945 
    taken with \textrm{Suzaku}.
    The top and second panels show those in  6.25--6.52~keV 
    and 6.5--9.0~keV, respectively, 
    taken with the four XIS sensors, 
    and re-binned with the orbital period of 5760~s.
    The third, fourth, and bottom panels shows 
    12--25~keV, 25--50~keV (HXD-PIN), and 52--114~keV (HXD-GSO)
    light curves, respectively.
    In these panels, each bin corresponds to a continuous 
    time interval without data gaps.
    Dotted line in the top, second and the fourth panels correspond 
    to 0.015~cts~s$^{-1}$, 0.027 and 0.075~cts~s$^{-1}$, respectively.
  }\label{fig:pin_lc_sub}
\end{figure}

Figure \ref{fig:pin_lc_sub} shows light curves of 
NGC~4945 taken with Suzaku in five energy ranges.
In the upper two panels (with the XIS), the data are binned 
at the orbital period (5760 sec), 
while in the other three panels (with the HXD) each time bin corresponds
to a continuous time interval which is free from any data gap.
The bin length in the latter case is typically 700--1500~sec, 
since HXD ``cleaned events'' suffer from two data gaps
per spacecraft orbit ($\sim$ 5760 sec),
mainly due to passages through high background regions
(at high magnetic longitudes and near the SAA).
We discarded data points shorter than 512~sec, 
since they have too large statistical errors.
The two light curves below 10~keV
are consistent with being constant. 
In the energy ranges above 10~keV, in contrast, we observed that
the source varied by a factor of $\sim 2$ within $\sim 20$~ks.
This agrees with the picture that the reflection from a distant cold 
material (plus possibly an extended thermal emission or scattering from 
an ionized medium) dominates the spectrum below 10~keV, 
while we are seeing the intrinsic AGN emission in the higher energy.

\begin{figure}
 \begin{center}
  \FigureFile(.9\linewidth,){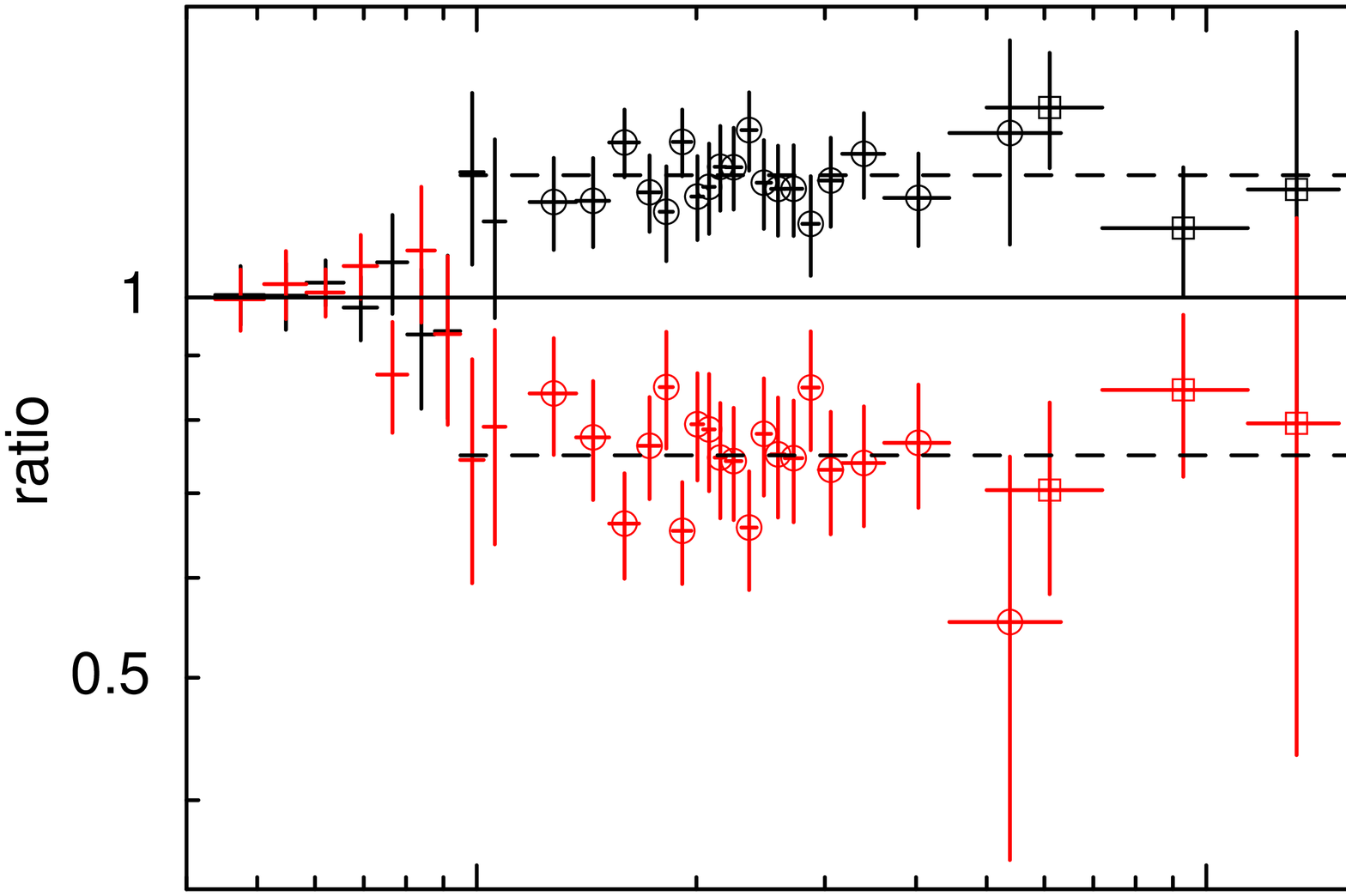}
 \end{center}
 \caption{Spectra from the ``high phase'' (black) and 
  ``low phase'' (red), 
  shown as their ratios to the time averaged spectrum.
  The XIS, HXD-PIN and HXD-GSO data points are shown 
  in crosses, circles,  and squares, respectively.
 }\label{fig:spec_ratio}
\end{figure}

In order to search the spectra for intensity-correlated changes,
we divided the entire observation period 
into two phases according to the 25--50~keV count rate;
``high flux'' and ``low flux'' phases, corresponding to periods 
when the count rate is higher and lower than 0.75 cts~s$^{-1}$, 
respectively. 
The net exposure is 33.7~ks for the high flux phase
and 31.7~ks for the low phase.
Figure \ref{fig:spec_ratio} shows the spectra of 
the two phases,
in the form of their ratios to the time-averaged spectrum.
Above 10~keV,  the flux of the high phase is
approximately twice as high as that of the low phase,
independent of energy.
In contrast, data points below 10~keV show no significant 
difference between the two phases, 
in agreement with the implication of the light curves.

We fitted each spectrum of the two phases
by the BC + scattered power-law model, and confirmed 
that the spectral parameters do not significantly change
as the flux varied, 
except for the normalization of the absorbed power-law component.

\begin{figure}
 \begin{center}
  \FigureFile(.9\linewidth,){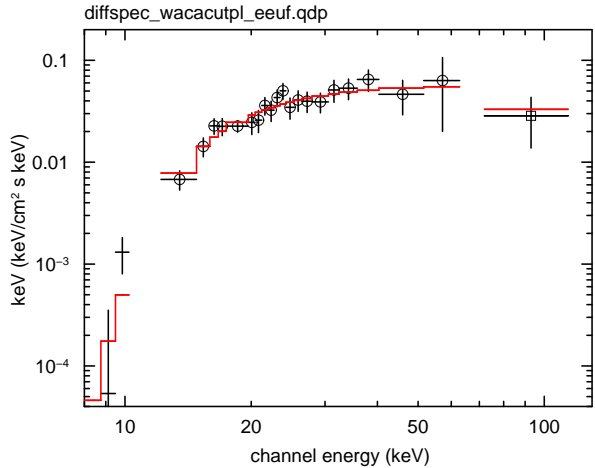}
 \end{center}
 \caption{Differential $\nu F \nu$ spectrum between 
 the two phases (black points), i.e. high minus low.
 on which an absorbed power-law model with $\Gamma = 1.6$ 
 is superposed (red line).
 }\label{fig:spec_diff}
\end{figure}

It is quite interesting to obtain the varying
spectral component, since it must be the pure intrinsic 
AGN emission without being diluted by any re-processed component.
Actually with the HXD, we were able to carry out 
this analysis in the present case.
We show in figure \ref{fig:spec_diff} the differential 
spectrum between the two phases,  
i.e. a spectrum of the high flux phase minus that of the low phase.
As shown there, an absorbed power-law reproduces the data 
successfully ($\chi^2$/d.o.f. = 141/216), with 
$\Gamma = 1.6^{+0.2}_{-0.3}$, 
$N_{\mathrm{H}} = 4.5^{+0.5}_{-1.0} \times 10^{24}$~cm$^{-2}$,
and $E_\mathrm{cut} \gtrsim 150$~keV (not constrained).
These parameter values agree with those we have derived from the 
time-averaged spectrum in the previous section. 
This confirms that the intrinsic component is quite dominant 
in the time-averaged spectrum, in the energy above $\sim 10$~keV,
and that the direct nuclear emission maintains its 
spectral shape as it varies.

%%------------------------------%%
\section{Discussion} 
%%------------------------------%%

\subsection{The Hard X-ray Emission\label{sec:disc1}}

In the hard X-ray ($\sim 4.5$--$120$~keV) band, the AGN emission
from NGC~4945 clearly exhibits at least two distinct spectral components.
One can be identified as the AGN primary emission 
transmitting through the Compton-thick absorber.
This component was observed above $\sim 10$~keV, showing rapid time 
variability during the observation. The spectrum is well reproduced 
by heavily absorbed power-law.
The other is dominant below $\sim 10$~keV, without showing
any time variability. The spectrum is characterized by a hard continuum
along with an edge structure and neutral/ionized fluorescence lines 
from heavy metals, suggesting its production  in re-processing materials
around the AGN.

The best-fit parameters of the primary power-law, 
$\Gamma = 1.6^{+0.1}_{-0.2}$ and 
$N_\mathrm{H} = 5.3^{+0.4}_{-0.9} \times 10^{24}$~cm$^{-2}$, 
agree with those derived by the previous hard X-ray missions 
(\cite{1993ApJ...409..155I}, \cite{2000ApJ...535L..87M}, 
\cite{2000A&A...356..463G}).
The absence of variability in the absorbing columns over several years
suggests the location of the absorber, 
namely, at a distance in excess of a light year from the nuclear 
continuum source.
The spectral index of $\sim 1.6$ lie at the relatively flat end
of the distribution among Seyfert galaxies  
(e.g. \cite{1997ApJ...488..164T}).
We cannot constrain the high energy spectral cutoff.
Although it is required in the broad band fitting around 200~keV 
(see table \ref{tab:first}),
when the systematic error on the GSO background modeling (2\%) 
is taken into account, we obtain only its lower limit of $\sim 80$~keV.
Nonetheless, a cutoff energy around 100-300~keV is consistent with 
the one  \citet{2000A&A...356..463G} obtained from this source,
and those measured in several Seyfert 1s 
(e.g. \cite{1995ApJ...438..672M}, \cite{1995ApJ...438L..63Z}, 
\cite{1999A&A...341L..27G}, \cite{1999A&A...346..407G},
\cite{1999A&A...351..937P}).

\subsection{The Absorbing Material}

The observed  average 10--50 keV flux was $1.6 \times 10^{-10}$ 
erg~s$^{-1}$~cm$^{-2}$, which is $\sim 1.6$ and $\sim 0.9$ times that 
measured by \textrm{RXTE} and \textrm{BeppoSAX}, respectively.
At the level of $N_\mathrm{H} \sim 4-5 \times 10^{24}$~cm$^{-2}$, 
the optical depth to electron scattering will become rather high, 
$\sim 3$ or so.
In such a case, the observed spectrum and flux 
would depend significantly on the geometry of the absorbing matter 
(\cite{1989MNRAS.236..603L}), even though the intrinsic luminosity 
is the same.
The model \texttt{cabs} assumes a configuration where 
a small blob of matter scatters 
some of the primary flux out of the line of sight.
If, on the contrary, the re-processing matter has a large solid angle 
to the central source, a fraction of the photons will also be scattered
into the line of sight, and the nuclear luminosity required to
explain the observed flux will decrease.
At the same time, such scattered photons would significantly 
dilute the time variability of the intrinsic source.
Based on the observed rapid (less than 1-day) hard X-ray 
variability in NGC~4945, 
\citet{2000ApJ...535L..87M} argued that 
the Compton-thick absorber must be geometrically-thin, 
with a half angle of less than 10$^{\circ}$ 
as seen from the central source.
If this is the case, the configuration assumed by 
\texttt{cabs} is considered to be more realistic.
%is rather plausible than the matter having a large solid angle.

While the present data strongly points to a geometrically 
thin absorber, the scenario requires a relatively high nuclear 
intrinsic luminosity. 
The extrapolation of the best-fit model to the 2--10 keV band, 
corrected for photo-electric absorption alone, 
yields a flux of $\sim 1.3 \times 10^{-10}$~erg~s$^{-1}$~cm$^{-2}$.
Further correcting this for the Thomson opacity, 
i.e. multiplying a factor $\exp(\tau_{\mathrm{es}})$ where
$\tau_{\mathrm{es}} = 1.21 N_\mathrm{H} \sigma_{\mathrm{Th}} $, 
we obtain $\sim 8 \times 10^{-9}$~erg~s$^{-1}$~cm$^{-2}$, 
and hence a 2--10~keV intrinsic luminosity of 
$\sim 1 \times 10^{43}$~erg~s$^{-1}$.
Using the bolometric corrections of \citet{2004MNRAS.351..169M}, 
a bolometric luminosity of $\sim 2 \times 10^{44}$~erg~s$^{-1}$ 
is derived, which in turn is $\sim 100$\% of the Eddington limit.
Although the inferred $L/L_\mathrm{Edd}$ is unplausibly large, 
this value is strongly dependent on $N_\mathrm{H}$, 
and there is also some scatter in the bolometric corrections 
(\cite{2004MNRAS.351..169M}).
Thus we can also get $L/L_\mathrm{Edd} \sim 30$\% within errors, 
and hence the scenario is energetically feasible.
%In addition, \cite{2000ApJ...535L..87M} implies that 
%at the level of $N_\matrhm{H} \sim$~4--5~$\times 10^{24}$~cm$^{-2}$, 
%a correction by \texttt{cabs} overestimates the intrinsic flux 
%by a factor of $\sim 3$, even when the absorber is geometrically-thin.
The derived bolometric luminosity is of the same order of magnitude 
as the observed infrared luminosity of the nucleus (\cite{Sanders2003}), 
and hence the AGN can dominate the energy output from the nucleus.

The flux of the intrinsic emission in the 12--50~keV band 
showed variations by a factor of 2 on a time scale of $\sim 20$~ks. 
Similar variability was observed with every previous mission
(\textrm{Ginga}, \textrm{BeppoSAX}, \textrm{RXTE}). 
From the 12--50~keV light curve, we calculated excess variance 
$\sigma^{2}_\mathrm{NXS}$ (\cite{1997ApJ...476...70N}) 
in the same manner  as \citet{2005MNRAS.358.1405O}, 
to find $\mathrm{log} (\sigma^{2}_\mathrm{NXS}) \sim -1.3 \pm 0.1$. 
This value lies well on the $M_\mathrm{BH}$ vs $\sigma^{2}_\mathrm{NXS}$
relationship derived from a number of Seyfert~1 AGNs;
the expected $\mathrm{log} (\sigma^{2}_\mathrm{NXS})$ for 
$M_\mathrm{BH} = 1.4 \times 10^6 M_{\odot}$ is -1.25
(\cite{2005MNRAS.358.1405O}).
This implies that the intrinsic variability is \textrm{not} diluted
by scattered photons, and supports the idea that 
the re-processing material be geometrically-thin.

\subsection{The Reflecting Materials}

The spectrum below 10~keV showed evidence of reflection 
from cold matter, and was modeled with a hard reflection 
continuum and neutral emission lines.
The derived equivalent width of the Fe~I~K$\alpha$ line,
ranging $\sim 1.3-1.6$~keV, is consistent with those expected 
when the line is produced in an optically-thick material 
with solar abundance (ranging between 1.3~keV and 2.7~keV,
\cite{1996MNRAS.280..823M}).
This suggests that the Fe K$\alpha$ I line and the reflection 
continuum are generated by the same material (see also 
\cite{2002MNRAS.335..241S}; \cite{2003ApJ...588..763D}).
We note that \citet{2000A&A...356..463G} preferred a transmission
origin for the line as the spectrum below 7~keV is not well fit by
reflection alone. 
However, this is due to there being additional complex
low energy emission from hot gas (some of which can be 
spatially resolved e.g. \cite{2003ApJ...588..763D}). 
Our detection of the compton shoulder at $f_{CS} \sim 0.10$ 
confirms a reflection origin, as transmission predicts
$f_{CS} \sim 0.3$ (\cite{2002MNRAS.337..147M}).  
The upper limit of $f_{CS}< 0.17$ also indicates that the reflector 
is probably seen nearly edge on (\cite{2002MNRAS.337..147M}).
We also detect for the first time the associated Fe~I~K$\beta$ and 
Ni~I~K$\alpha$ lines from the reflector. 
The flux of these lines are $10^{+3}_{-3}~\%$ and $10^{+5}_{-5}~\%$ 
of that of Fe~I~K$\alpha$, respectively, 
which also agree with the value expected from the neutral reflection.

%This suggests that the Fe~I~line and the reflection continuum 
%are generated by the same material.
%Note that \cite{2000A&A...356..463G} inferred that the line is
%produced in transmission by the absorbing medium, 
%based on lack of a reflection continuum in their spectrum.
%The other recent works, in agreement with ours,
%(\cite{2002MNRAS.335..241S}, \cite{2003ApJ...588..763D})
%certainly observed the hard reflection continuum below 10~keV.
%The Fe~I~K$\beta$ and Ni~I~K$\beta$ lines have been detected for 
%the first time from this source.

%\red{
%In addition, if the line is produced in transmission 
%through near-spherical absorbing medium
%the Compton shoulder parameter $f_\mathrm{CS}$ would be 
%$> 0.3$ (\cite{2002MNRAS.337..147M}).
%In contrast, our observed value of $f_\mathrm{CS}$ ($< 0.17$)
%suggests that the line is originating in the reflection, 
%and its smallness prefers the reflector being located 
%at a near edge-on angle (\cite{2002MNRAS.337..147M}).
%}

The obtained $R$ is quite small at a~few~$\times 10^{-3}$. 
As mentioned in \S\ref{sec:disc1}, 
the estimation of the intrinsic flux, and hence of $R$, 
is strongly affected by the geometry of the Compton-thick absorber.
In the most extreme case, 
i.e. when the absorber is spherically surrounding the continuum source,
the intrinsic flux would be 3--6 times lower than derived here
(\cite{1999NewA....4..191M}, \cite{2000ApJ...535L..87M}).
Even in that case, $R$ still remains $\lesssim$~a~few~$\times 10^{-2}$,
which is significantly smaller than those of the other Compton-thick 
Seyfert 2 AGNs; e.g. $R = 1.2-1.4$ in MKN~3
(\cite{2005MNRAS.360.1123P}, Awaki et al. 2007). 
This might be explained by assuming the reflector to be located nearly 
edge-on angle; so we fitted the spectrum with $i = 85^\circ$, 
to find that the $R$ value increases by only a factor of $\sim 2$. 
The small $R$ implies that we see in the reflection 
only a small part of the illuminated disk 
on the far side of itself.
%rather than a putative torus with a large covering factor.

All the above evidence on the re-processed signals implies
that the geometrically-thin/disk-like reflector is viewed 
from a near edge-on angle. 
Considering that the H$_2$O megamaser emission requires
an edge-on inner disk geometry in this source 
(\cite{1997ApJ...481L..23G}), the simplest explanation of the 
reflector, absorber and maser source is that these are all 
the same structure, namely the accretion disk
(\cite{2000ApJ...535L..87M}, \cite{2006ApJ...636...75M}).
Thus the accretion structure in this source could be 
different from a typical AGN which is considered to have 
a putative torus with a large covering factor.

%Another remaining question is the origin of the ionized Fe line
%observed at 6.683$^{+0.015}_{-0.015}$~keV.
%The line energy is consistent with either a resonance (6.700~keV) or
%an inter-combination (6.675~keV) line, or their mixture.
%Although BC + Scattered Continuum model gives better $\chi^2$ 
%than BC + Thermal Plasma model, this is likely due to the fact 
%that the former is more flexible than the latter.
%Thus, we just discuss physical plausibily
%of the spectral parameters derived from each model.

\subsection{The Ionized Iron Line --Consistency Check--}

Another remaining question is the origin of the ionized Fe line 
observed at $6.683^{+0.015}_{-0.015}$~keV. 
The BC + Scattered power-law model gives a slightly better fit 
than the BC + thermal model, to the spectral curvature around 4~keV.
However, this is probably simply due to there being more free 
parameters in this model, and it is thus difficult to provide 
any clearcut preference on the origin of the soft X-ray continuum
and the ionized line from our data.
In this section, we therefore only discuss physical consistency 
of each of our scenarios.

From its energy, the ionized line is consistent with the resonance line,
$w$, at 6.700~keV or the intercombination lines, $x+y$, 
at 6.682 and 6.668~keV, or their mixture. 
A strong contribution from the forbidden line, $z$, at 6.637~keV, 
is clearly ruled out by the mean line energy. 
Collisionally ionized plasmas have $G \equiv (x+y+z)/w \sim 1$
while a pure photo-ionized plasma has $G>1$ as the resonance line is weak.
Note that contamination from dielectronic lines is not important at the
inferred temperature of $\sim 7$~keV (\cite{2001MNRAS.327L..42O}).
However, continuum photons enhance $w$ through resonance line
scattering, so the observed line energy can also be consistent with
photo-ionization (\cite{2005MNRAS.357..599B}). 
%The continuum shape is also equally ambiguous. 

The temperature of $6-8$~keV estimated from the thermal plasma model 
is rather high compared to those expected 
from ordinary star-burst activity ($\sim 500$~eV), 
and difficult to produce 
(\cite{1994ApJ...430..511S}, \cite{2000MNRAS.314..511S}).
However, a similar temperature was observed in a nearby 
pure star-burst galaxy NGC~253 (\cite{2001A&A...365L.174P}).
From the scattered continuum model, we obtained 
$f_\mathrm{scat} \sim \mathrm{several}~\times 10^{-4}$.
The column density of the photo-ionized gas
can be estimated on the basis of this ratio.
In order to have an order of magnitude estimate
we assume a covering factor of $\sim 0.1$, 
then the gas column density would be $10^{21}-10^{22}$~cm$^{-2}$.
This value lies within those estimated in other obscured AGNs 
(e.g. \cite{2006A&A...448..499B}).
The equivalent width of the Fe~XXV~K$\alpha$ and 
Fe~XXVI~K$\alpha$ lines with respect to the scattered power-law, 
i.e. $\sim 0.5$~keV and $< 0.1$~keV, respectively, 
are consistent with the both components originating 
from a photo-ionized gas 
(see calculations in \cite{2002A&A...387...76B};
\cite{2005MNRAS.357..599B}).
To highly ionize the iron, the region where the line originates 
needs to be close to the central source so that its flux 
could potentially vary as the primary continuum does.
We searched in vain for any time variability of the ionized lines.

Therefore, both of our scenarios are considered to be physically 
consistent. Yet another possibility is that the ionized line, 
together with the modeled scattered continuum, 
arise from unresolved point sources (\cite{2003ApJ...588..763D}).
%Thus the origin of the ionized Fe-K line and the continuum 
%remains unsettled.

%%------------------------------%%
\vspace*{2ex}
\noindent
{\bf Acknowledgments} \\[1ex]
%%------------------------------%%
The authors would like to express their thanks to the Suzaku 
team members, 
and the anonymous referee for the comments to improve the paper.
TI and PG, being research fellows of the Japan Society 
for the Promotion of Science, thank the Society.
CD thanks the ISAS/JAXA visiting professor program for support.
GM acknowledges the support by the US Department of Energy 
contract to SLAC no. DE-AC3-76SF00515, 
and NASA Suzaku grant no. NNX07AB05G.
The authors thank Dr. Andrzej Zdziarski for his helpful discussions.


\vspace*{2ex}

\begin{thebibliography}{61}
\expandafter\ifx\csname natexlab\endcsname\relax\def\natexlab#1{#1}\fi
\expandafter\ifx\csname href\endcsname\relax
  \def\href#1#2{}\fi
\expandafter\ifx\csname urllinklabel\endcsname\relax
  \def\urllinklabel{[LINK]}\fi
\expandafter\ifx\csname adsurllinklabel\endcsname\relax
  \def\adsurllinklabel{[ADS]}\fi

\bibitem[{{Antonucci} \& {Miller}(1985)}]{1985ApJ...297..621A}
{Antonucci}, R.~R.~J. \& {Miller}, J.~S. 1985, \apj, 297, 621


\bibitem[{{Awaki} \& {Koyama}(1993)}]{1993AdSpR..13..221A}
{Awaki}, H. \& {Koyama}, K. 1993, Advances in Space Research, 13, 221


\bibitem[{{Awaki} {et~al.}(1990){Awaki}, {Koyama}, {Kunieda}, \&
  {Tawara}}]{1990Natur.346..544A}
{Awaki}, H., {Koyama}, K., {Kunieda}, H., \& {Tawara}, Y. 1990, \nat, 346, 544

\bibitem[{{Bianchi} {et~al.}(2006){Bianchi}, {Guainazzi}, \&
  {Chiaberge}}]{2006A&A...448..499B}
{Bianchi}, S., {Guainazzi}, M., \& {Chiaberge}, M. 2006, \aap, 448, 499

\bibitem[{{Bianchi} \& {Matt}(2002)}]{2002A&A...387...76B}
{Bianchi}, S. \& {Matt}, G. 2002, \aap, 387, 76


\bibitem[{{Bianchi} {et~al.}(2005{\natexlab{a}}){Bianchi}, {Matt}, {Nicastro},
  {Porquet}, \& {Dubau}}]{2005MNRAS.357..599B}
{Bianchi}, S., {Matt}, G., {Nicastro}, F., {Porquet}, D., \& {Dubau}, J.
  2005{\natexlab{a}}, \mnras, 357, 599


\bibitem[{{Bianchi} {et~al.}(2005{\natexlab{b}}){Bianchi}, {Miniutti},
  {Fabian}, \& {Iwasawa}}]{2005MNRAS.360..380B}
{Bianchi}, S., {Miniutti}, G., {Fabian}, A.~C., \& {Iwasawa}, K.
  2005{\natexlab{b}}, \mnras, 360, 380

\bibitem[{{Boldt}(1987)}]{1987PhR...146..215B}
{Boldt}, E. 1987, \physrep, 146, 215


\bibitem[{{Chen} \& {Huang}(1997)}]{1997ApJ...479L..23C}
{Chen}, Y. \& {Huang}, J.-H. 1997, \apjl, 479, L23+


\bibitem[{{Churazov} {et~al.}(2007)}]{2007A&A...467..529C}
{Churazov}, E. et~al. 2007, \aap, 467, 529


\bibitem[{{Comastri} {et~al.}(1995){Comastri}, {Setti}, {Zamorani}, \&
  {Hasinger}}]{1995A&A...296....1C}
{Comastri}, A., {Setti}, G., {Zamorani}, G., \& {Hasinger}, G. 1995, \aap, 296,
  1


\bibitem[{{Done} {et~al.}(1996){Done}, {Madejski}, \&
  {Smith}}]{1996ApJ...463L..63D}
{Done}, C., {Madejski}, G.~M., \& {Smith}, D.~A. 1996, \apjl, 463, L63+


\bibitem[{{Done} {et~al.}(2003){Done}, {Madejski}, {{\.Z}ycki}, \&
  {Greenhill}}]{2003ApJ...588..763D}
{Done}, C., {Madejski}, G.~M., {{\.Z}ycki}, P.~T., \& {Greenhill}, L.~J. 2003,
  \apj, 588, 763


\bibitem[{{Gandhi} \& {Fabian}(2003)}]{2003MNRAS.339.1095G}
{Gandhi}, P. \& {Fabian}, A.~C. 2003, \mnras, 339, 1095


\bibitem[{{Gilli} {et~al.}(2007){Gilli}, {Comastri}, \&
  {Hasinger}}]{2007A&A...463...79G}
{Gilli}, R., {Comastri}, A., \& {Hasinger}, G. 2007, \aap, 463, 79


\bibitem[{{Greenhill} {et~al.}(1997){Greenhill}, {Moran}, \&
  {Herrnstein}}]{1997ApJ...481L..23G}
{Greenhill}, L.~J., {Moran}, J.~M., \& {Herrnstein}, J.~R. 1997, \apjl, 481,
  L23+


\bibitem[{{Gruber} {et~al.}(1999){Gruber}, {Matteson}, {Peterson}, \&
  {Jung}}]{1999ApJ...520..124G}
{Gruber}, D.~E., {Matteson}, J.~L., {Peterson}, L.~E., \& {Jung}, G.~V. 1999,
  \apj, 520, 124


\bibitem[{{Guainazzi} \& {Bianchi}(2007)}]{2007MNRAS.374.1290G}
{Guainazzi}, M. \& {Bianchi}, S. 2007, \mnras, 374, 1290


\bibitem[{{Guainazzi} {et~al.}(2000){Guainazzi}, {Matt}, {Brandt}, {Antonelli},
  {Barr}, \& {Bassani}}]{2000A&A...356..463G}
{Guainazzi}, M., {Matt}, G., {Brandt}, W.~N., {Antonelli}, L.~A., {Barr}, P.,
  \& {Bassani}, L. 2000, \aap, 356, 463


\bibitem[{{Guainazzi} {et~al.}(1999{\natexlab{a}}){Guainazzi}, {Matt},
  {Molendi}, {Orr}, {Fiore}, {Grandi}, {Matteuzzi}, {Mineo}, {Perola},
  {Parmar}, \& {Piro}}]{1999A&A...341L..27G}
{Guainazzi}, M., {Matt}, G., {Molendi}, S., {Orr}, A., {Fiore}, F., {Grandi},
  P., {Matteuzzi}, A., {Mineo}, T., {Perola}, G.~C., {Parmar}, A.~N., \&
  {Piro}, L. 1999{\natexlab{a}}, \aap, 341, L27


\bibitem[{{Guainazzi} {et~al.}(1999{\natexlab{b}}){Guainazzi}, {Perola},
  {Matt}, {Nicastro}, {Bassani}, {Fiore}, {dal Fiume}, \&
  {Piro}}]{1999A&A...346..407G}
{Guainazzi}, M., {Perola}, G.~C., {Matt}, G., {Nicastro}, F., {Bassani}, L.,
  {Fiore}, F., {dal Fiume}, D., \& {Piro}, L. 1999{\natexlab{b}}, \aap, 346,
  407


\bibitem[{{Heiles} \& {Cleary}(1979)}]{1979AuJPA..47....1H}
{Heiles}, C. \& {Cleary}, M.~N. 1979, Australian Journal of Physics
  Astrophysical Supplement, 47, 1


\bibitem[{{Ikebe} {et~al.}(2000){Ikebe}, {Leighly}, {Tanaka}, {Nakagawa},
  {Terashima}, \& {Komossa}}]{2000MNRAS.316..433I}
{Ikebe}, Y., {Leighly}, K., {Tanaka}, Y., {Nakagawa}, T., {Terashima}, Y., \&
  {Komossa}, S. 2000, \mnras, 316, 433


\bibitem[{{Illarionov} {et~al.}(1979){Illarionov}, {Kallman}, {McCray}, \&
  {Ross}}]{1979ApJ...228..279I}
{Illarionov}, A., {Kallman}, T., {McCray}, R., \& {Ross}, R. 1979, \apj, 228,
  279


\bibitem[{{Ishisaki}(2007)}]{2007PASJ...59S.113I}
{Ishisaki}, Y.~et.~al. 2007, \pasj, 59, 113


\bibitem[{{Isobe} {et~al.}(2007){Isobe}, {Kubota}, {Makishima}, {Gandhi},
  {Griffiths}, {Dewangan}, {Itoh}, \& {Mizuno}}]{Isobe2007}
{Isobe}, N., {Kubota}, A., {Makishima}, K., {Gandhi}, P., {Griffiths}, R.,
  {Dewangan}, G.~C., {Itoh}, T., \& {Mizuno}, T. 2007, \pasj, accepted


\bibitem[{{Iwasawa} {et~al.}(1993){Iwasawa}, {Koyama}, {Awaki}, {Kunieda},
  {Makishima}, {Tsuru}, {Ohashi}, \& {Nakai}}]{1993ApJ...409..155I}
{Iwasawa}, K., {Koyama}, K., {Awaki}, H., {Kunieda}, H., {Makishima}, K.,
  {Tsuru}, T., {Ohashi}, T., \& {Nakai}, N. 1993, \apj, 409, 155


\bibitem[{{Iwasawa} {et~al.}(1994){Iwasawa}, {Yaqoob}, {Awaki}, \&
  {Ogasaka}}]{1994PASJ...46L.167I}
{Iwasawa}, K., {Yaqoob}, T., {Awaki}, H., \& {Ogasaka}, Y. 1994, \pasj, 46,
  L167


\bibitem[{{Kinkhabwala} {et~al.}(2002){Kinkhabwala}, {Sako}, {Behar}, {Kahn},
  {Paerels}, {Brinkman}, {Kaastra}, {Gu}, \& {Liedahl}}]{2002ApJ...575..732K}
{Kinkhabwala}, A., {Sako}, M., {Behar}, E., {Kahn}, S.~M., {Paerels}, F.,
  {Brinkman}, A.~C., {Kaastra}, J.~S., {Gu}, M.~F., \& {Liedahl}, D.~A. 2002,
  \apj, 575, 732


\bibitem[{{Kokubun} {et~al.}(2007)}]{2007PASJ...59S..53K}
{Kokubun}, M.~et~al. 2007, \pasj, 59, 53


\bibitem[{{Koyama} {et~al.}(1989){Koyama}, {Inoue}, {Tanaka}, {Awaki},
  {Takano}, {Ohashi}, \& {Matsuoka}}]{1989PASJ...41..731K}
{Koyama}, K., {Inoue}, H., {Tanaka}, Y., {Awaki}, H., {Takano}, S., {Ohashi},
  T., \& {Matsuoka}, M. 1989, \pasj, 41, 731


\bibitem[{{Koyama} {et~al.}(2007)}]{2007PASJ...59S..23K}
{Koyama}, K.~et~al. 2007, \pasj, 59, 23


\bibitem[{{Leahy} {et~al.}(1989){Leahy}, {Matsuoka}, {Kawai}, \&
  {Makino}}]{1989MNRAS.236..603L}
{Leahy}, D.~A., {Matsuoka}, M., {Kawai}, N., \& {Makino}, F. 1989, \mnras, 236,
  603


\bibitem[{{Madau} {et~al.}(1994){Madau}, {Ghisellini}, \&
  {Fabian}}]{1994MNRAS.270L..17M}
{Madau}, P., {Ghisellini}, G., \& {Fabian}, A.~C. 1994, \mnras, 270, L17+


\bibitem[{{Madejski} {et~al.}(2006){Madejski}, {Done}, {{\.Z}ycki}, \&
  {Greenhill}}]{2006ApJ...636...75M}
{Madejski}, G., {Done}, C., {{\.Z}ycki}, P.~T., \& {Greenhill}, L. 2006, \apj,
  636, 75


\bibitem[{{Madejski} {et~al.}(2000){Madejski}, {{\.Z}ycki}, {Done}, {Valinia},
  {Blanco}, {Rothschild}, \& {Turek}}]{2000ApJ...535L..87M}
{Madejski}, G., {{\.Z}ycki}, P., {Done}, C., {Valinia}, A., {Blanco}, P.,
  {Rothschild}, R., \& {Turek}, B. 2000, \apjl, 535, L87


\bibitem[{{Madejski} {et~al.}(1995){Madejski}, {Zdziarski}, {Turner}, {Done},
  {Mushotzky}, {Hartman}, {Gehrels}, {Connors}, {Fabian}, {Nandra}, {Celotti},
  {Rees}, {Johnson}, {Grove}, \& {Starr}}]{1995ApJ...438..672M}
{Madejski}, G.~M.~et~al. 1995, \apj, 438, 672


\bibitem[{{Magdziarz} \& {Zdziarski}(1995)}]{1995MNRAS.273..837M}
{Magdziarz}, P. \& {Zdziarski}, A.~A. 1995, \mnras, 273, 837


\bibitem[{{Malizia} {et~al.}(2003){Malizia}, {Bassani}, {Stephen}, {Di Cocco},
  {Fiore}, \& {Dean}}]{2003ApJ...589L..17M}
{Malizia}, A., {Bassani}, L., {Stephen}, J.~B., {Di Cocco}, G., {Fiore}, F., \&
  {Dean}, A.~J. 2003, \apjl, 589, L17


\bibitem[{{Marconi} {et~al.}(2004){Marconi}, {Risaliti}, {Gilli}, {Hunt},
  {Maiolino}, \& {Salvati}}]{2004MNRAS.351..169M}
{Marconi}, A., {Risaliti}, G., {Gilli}, R., {Hunt}, L.~K., {Maiolino}, R., \&
  {Salvati}, M. 2004, \mnras, 351, 169


\bibitem[{{Matt}(2002)}]{2002MNRAS.337..147M}
{Matt}, G. 2002, \mnras, 337, 147


\bibitem[{{Matt} {et~al.}(1996){Matt}, {Brandt}, \&
  {Fabian}}]{1996MNRAS.280..823M}
{Matt}, G., {Brandt}, W.~N., \& {Fabian}, A.~C. 1996, \mnras, 280, 823


\bibitem[{{Matt} {et~al.}(2000){Matt}, {Fabian}, {Guainazzi}, {Iwasawa},
  {Bassani}, \& {Malaguti}}]{2000MNRAS.318..173M}
{Matt}, G., {Fabian}, A.~C., {Guainazzi}, M., {Iwasawa}, K., {Bassani}, L., \&
  {Malaguti}, G. 2000, \mnras, 318, 173


\bibitem[{{Matt} {et~al.}(1997){Matt}, {Guainazzi}, {Frontera}, {Bassani},
  {Brandt}, {Fabian}, {Fiore}, {Haardt}, {Iwasawa}, {Maiolino}, {Malaguti},
  {Marconi}, {Matteuzzi}, {Molendi}, {Perola}, {Piraino}, \&
  {Piro}}]{1997A&A...325L..13M}
{Matt}, G.~et~al. 1997, \aap, 325, L13


\bibitem[{{Matt} {et~al.}(1999){Matt}, {Pompilio}, \& {La
  Franca}}]{1999NewA....4..191M}
{Matt}, G., {Pompilio}, F., \& {La Franca}, F. 1999, New Astronomy, 4, 191


\bibitem[{{Mauersberger} {et~al.}(1996){Mauersberger}, {Henkel}, {Whiteoak},
  {Chin}, \& {Tieftrunk}}]{1996A&A...309..705M}
{Mauersberger}, R., {Henkel}, C., {Whiteoak}, J.~B., {Chin}, Y.-N., \&
  {Tieftrunk}, A.~R. 1996, \aap, 309, 705


\bibitem[{{Mitsuda} {et~al.}(2007)}]{2007PASJ...59S...1M}
{Mitsuda}, K.~et~al. 2007, \pasj, 59, 1


\bibitem[{{Moorwood} {et~al.}(1996){Moorwood}, {van der Werf}, {Kotilainen},
  {Marconi}, \& {Oliva}}]{1996A&A...308L...1M}
{Moorwood}, A.~F.~M., {van der Werf}, P.~P., {Kotilainen}, J.~K., {Marconi},
  A., \& {Oliva}, E. 1996, \aap, 308, L1+


\bibitem[{{Nandra} {et~al.}(1997){Nandra}, {George}, {Mushotzky}, {Turner}, \&
  {Yaqoob}}]{1997ApJ...476...70N}
{Nandra}, K., {George}, I.~M., {Mushotzky}, R.~F., {Turner}, T.~J., \&
  {Yaqoob}, T. 1997, \apj, 476, 70

\bibitem[{{Oelgoetz} \& {Pradhan}(2001)}]{2001MNRAS.327L..42O}
{Oelgoetz}, J. \& {Pradhan}, A.~K. 2001, \mnras, 327, L42

\bibitem[{{O'Neill} {et~al.}(2005){O'Neill}, {Nandra}, {Papadakis}, \&
  {Turner}}]{2005MNRAS.358.1405O}
{O'Neill}, P.~M., {Nandra}, K., {Papadakis}, I.~E., \& {Turner}, T.~J. 2005,
  \mnras, 358, 1405


\bibitem[{{Ott} {et~al.}(2001){Ott}, {Whiteoak}, {Henkel}, \&
  {Wielebinski}}]{2001A&A...372..463O}
{Ott}, M., {Whiteoak}, J.~B., {Henkel}, C., \& {Wielebinski}, R. 2001, \aap,
  372, 463


\bibitem[{{Perola} {et~al.}(1999){Perola}, {Matt}, {Cappi}, {Dal Fiume},
  {Fiore}, {Guainazzi}, {Mineo}, {Molendi}, {Nicastro}, {Piro}, \&
  {Stirpe}}]{1999A&A...351..937P}
{Perola}, G.~C., {Matt}, G., {Cappi}, M., {Dal Fiume}, D., {Fiore}, F.,
  {Guainazzi}, M., {Mineo}, T., {Molendi}, S., {Nicastro}, F., {Piro}, L., \&
  {Stirpe}, G. 1999, \aap, 351, 937


\bibitem[{{Pietsch} {et~al.}(2001){Pietsch}, {Roberts}, {Sako}, {Freyberg},
  {Read}, {Borozdin}, {Branduardi-Raymont}, {Cappi}, {Ehle}, {Ferrando},
  {Kahn}, {Ponman}, {Ptak}, {Shirey}, \& {Ward}}]{2001A&A...365L.174P}
{Pietsch}, W. et~al. 2001, \aap, 365, L174

\bibitem[{{Pounds} \& {Vaughan}(2006)}]{2006MNRAS.368..707P}
{Pounds}, K. \& {Vaughan}, S. 2006, \mnras, 368, 707


\bibitem[{{Pounds} \& {Page}(2005)}]{2005MNRAS.360.1123P}
{Pounds}, K.~A. \& {Page}, K.~L. 2005, \mnras, 360, 1123

\bibitem[{{Sanders} {et~al.}(2003) }]{Sanders2003}
{Sanders}, D.~B., {Mazzarella}, J.~M., {Kim}, D.~-C \&
{Surace}, J.~A. 2003, \apj, 126, 1607

\bibitem[{{Revnivtsev} {et~al.}(2003){Revnivtsev}, {Gilfanov}, {Sunyaev},
  {Jahoda}, \& {Markwardt}}]{2003A&A...411..329R}
{Revnivtsev}, M., {Gilfanov}, M., {Sunyaev}, R., {Jahoda}, K., \& {Markwardt},
  C. 2003, \aap, 411, 329


\bibitem[{{Schurch} {et~al.}(2002){Schurch}, {Roberts}, \&
  {Warwick}}]{2002MNRAS.335..241S}
{Schurch}, N.~J., {Roberts}, T.~P., \& {Warwick}, R.~S. 2002, \mnras, 335, 241


\bibitem[{{Strickland} \& {Stevens}(2000)}]{2000MNRAS.314..511S}
{Strickland}, D.~K. \& {Stevens}, I.~R. 2000, \mnras, 314, 511


\bibitem[{{Suchkov} {et~al.}(1994){Suchkov}, {Balsara}, {Heckman}, \&
  {Leitherner}}]{1994ApJ...430..511S}
{Suchkov}, A.~A., {Balsara}, D.~S., {Heckman}, T.~M., \& {Leitherner}, C. 1994,
  \apj, 430, 511


\bibitem[{{Takahashi} {et~al.}(2007)}]{Takahashi2007}
{Takahashi}, H.~et~al. \pasj, submitted.


\bibitem[{{Takahashi} {et~al.}(2007)}]{2007PASJ...59S..35T}
{Takahashi}, T.~et~al. 2007, \pasj, 59, 35


\bibitem[{{Turner} {et~al.}(1997){Turner}, {George}, {Nandra}, \&
  {Mushotzky}}]{1997ApJ...488..164T}
{Turner}, T.~J., {George}, I.~M., {Nandra}, K., \& {Mushotzky}, R.~F. 1997,
  \apj, 488, 164


\bibitem[{{Ueno} {et~al.}(1994{\natexlab{a}}){Ueno}, {Koyama}, {Nishida},
  {Yamauchi}, \& {Ward}}]{1994ApJ...431L...1U}
{Ueno}, S., {Koyama}, K., {Nishida}, M., {Yamauchi}, S., \& {Ward}, M.~J.
  1994{\natexlab{a}}, \apjl, 431, L1


\bibitem[{{Ueno} {et~al.}(1994{\natexlab{b}}){Ueno}, {Mushotzky}, {Koyama},
  {Iwasawa}, {Awaki}, \& {Hayashi}}]{1994PASJ...46L..71U}
{Ueno}, S., {Mushotzky}, R.~F., {Koyama}, K., {Iwasawa}, K., {Awaki}, H., \&
  {Hayashi}, I. 1994{\natexlab{b}}, \pasj, 46, L71


\bibitem[{{Young} {et~al.}(2001){Young}, {Wilson}, \&
  {Shopbell}}]{2001ApJ...556....6Y}
{Young}, A.~J., {Wilson}, A.~S., \& {Shopbell}, P.~L. 2001, \apj, 556, 6


\bibitem[{{Zdziarski} {et~al.}(1995){Zdziarski}, {Johnson}, {Done}, {Smith}, \&
  {McNaron-Brown}}]{1995ApJ...438L..63Z}
{Zdziarski}, A.~A., {Johnson}, W.~N., {Done}, C., {Smith}, D., \&
  {McNaron-Brown}, K. 1995, \apjl, 438, L63


\end{thebibliography}
\end{document}